\documentclass[reprint, aps, prd, letterpaper, noshowpacs, amsmath, %
amssymb, amsfonts, nofootinbib, floatfix, superscriptaddress, %
twoside]{revtex4-1}
\pdfoutput=1

\usepackage[T3,T1]{fontenc}
\usepackage{mathrsfs} % For \mathscr fonts
\usepackage{bm} % For \bm fonts
\makeatletter
\@ifclassloaded{beamer}
  { % if this is a beamer document, use sans-serif fonts
    \typeout{UsePackages: Detected beamer}
    \usepackage{tgheros}
    
  }
  { % otherwise:
    \typeout{UsePackages: Did not detect beamer}
    \ifx\asybeamer\undefined % use times for articles
    \typeout{UsePackages: Detected article}
    \usepackage[varg]{txfonts} % For math
    \usepackage{tgtermes} % For text
    \else % use beamer fonts for asy documents with beamer
    \typeout{Fonts: Detected asy for beamer}
    \usepackage{tgheros}
    
    \fi
  }
\usepackage{microtype} % For nicer alignment of text
\def\MT@register@subst@font{
  \MT@exp@one@n\MT@in@clist\font@name\MT@font@list
  \ifMT@inlist@\else\xdef\MT@font@list{\MT@font@list\font@name,}\fi}
\makeatother

\usepackage{graphicx} %
\usepackage[x11names,svgnames,rgb]{xcolor} %
\usepackage{xspace}
\usepackage{braket}
\usepackage{accents}
\usepackage{siunitx}
\usepackage{mathtools}
\DeclareSymbolFontAlphabet{\mathrm}{operators}

\definecolor{CiteColor}{rgb}{0.18039, 0.18824, 0.57255}
\definecolor{UrlColor} {rgb}{0.741, 0.173, 0.000}
\definecolor{DarkUrlColor} {rgb}{0.500, 0.110, 0.000}
\definecolor{LinkColor}{rgb}{0.25098, 0.47843, 0.04706}
\makeatletter %
\newcommand{\ShowFont}{%
  \typeout{The main font is \f@encoding \space \f@family \space %
    \f@series \space \f@shape \space at \f@size pt.}%
  \typeout{The math font sizes are \tf@size pt (main), \sf@size pt %
    (script), and \ssf@size pt (scriptscript).}%
  \typeout{The linewidth is \the\linewidth}} %
\makeatother %
\makeatletter

\def\@seccntformat#1{\csname the#1\endcsname.~}%

\def\section{%
  \@startsection {section}
  {1} {\z@} {0.55cm \@plus1ex \@minus .02ex}%
    {0.225cm} { \normalfont\bfseries \centering}%
}%
\def\subsection{%
  \@startsection {subsection}
  {2} {\z@ } {0.45cm \@plus 0.8ex \@minus 0.2ex}%
  {0.1125cm}{\normalfont \bfseries \centering }}
\def\subsubsection{%
  \@startsection {subsubsection}
  {3} {\z@ } {0.4cm \@plus 0.6ex \@minus 0.1ex}%
  {0.075cm}{\normalfont \it \centering }}

\newcommand{\surnames}[1]{\def\@surnamelist{#1}\relax}
\surnames{\ }
\ifx \@shorttitle \@empty
\else
  \def\@oddhead{\small \MakeUppercase{\@shorttitle} \hfill}
\fi
\ifx \@surnamelist \@empty
\else
  \def\@evenhead{\small \MakeUppercase{\@surnamelist} \hfill}
\fi
\def\@oddfoot{\reset@font\hfil\thepage\hfil}%
\def\@evenfoot{\reset@font\hfil\thepage\hfil}%
\g@addto@macro\maketitle{\global\@specialpagetrue\gdef\@specialstyle{plain}}

\usepackage{xpatch}% http://ctan.org/pkg/xpatch
\makeatletter
\xpatchcmd{\@ssect@ltx}{\@xsect}{\edef\@currentlabelname{#8}\@xsect}{}{}% Patch \<section>*
\xpatchcmd{\@sect@ltx}{\@xsect}{\edef\@currentlabelname{#8}\@xsect}{}{}% Patch \<section>
\makeatother

\usepackage[colorlinks, plainpages=false,
hyperfigures=true]{hyperref}
\hypersetup{linkcolor=LinkColor}
\hypersetup{citecolor=CiteColor}
\hypersetup{urlcolor=UrlColor}
\hypersetup{setpagesize=false}
\hypersetup{pdfborder=0 0 0}
\makeatother

\let\Originalddefinition\d
\renewcommand{\d}{\ensuremath{\mathrm{d}}}

\newcommand{\e}{\ensuremath{\mathrm{e}}}
\let\Originalidefinition\i
\renewcommand{\i}{\ensuremath{\mathrm{i}}}

\newcommand{\scriplus}{\ensuremath{\mathscr{I}^{+}}}

\let\Originalcdefinition\c
\renewcommand{\c}{\mathrm{c}}

\newcommand{\MSun}{\ensuremath{M_\odot}\xspace}

\DeclareSIUnit{\strain}{strain}
\DeclareSIPrePower{\root}{1/2}
\DeclareSIUnit{\parsec}{pc}
\DeclareSIUnit{\yr}{yr}
\DeclareSIUnit{\year}{yr}
\DeclareSIUnit{\lightyear}{ly}
\DeclareSIUnit{\SolarMass}{\ensuremath{\MSun}}
\DeclareSIUnit{\Mass}{\ensuremath{M}}
\newcommand{\abs} [1]{\left\lvert{#1}\right\rvert}

\newcommand{\co}[1]{\ensuremath{\bar{#1}}}

\newcommand{\defined}{\coloneqq}

\newcommand{\roughly}{\mathord{\sim}} % Different from \sim in spacing
\newcommand{\D}{\ensuremath{\mathfrak{D}} }

\newcommand{\sYlm}[1]{\ensuremath{\scripts{_{s}}{Y}{_{#1}}}}
\newcommand{\sYbarlm}[1]{\ensuremath{\scripts{_{s}}{\bar{Y}}{_{#1}}}}
\newcommand{\mTwoYlm}[1]{\ensuremath{\scripts{_{-2}}{Y}{_{#1}}}}
\newcommand{\mTwoYbarlm}[1]{\ensuremath{\scripts{_{-2}}{\bar{Y}}{_{#1}}}}
\DeclareSymbolFont{tipa}{T3}{tipa}{m}{n}
\DeclareMathAccent{\ibreve}{\mathalpha}{tipa}{'020}
\newcommand{\Rotated}[1]{\ensuremath{\ibreve{#1}}}

\newcommand{\h}{\ensuremath{h} }
\newcommand{\hp}{\ensuremath{h_{+}} }
\newcommand{\hc}{\ensuremath{h_{\times}} }

\makeatletter
\newcommand{\foreign}[1]{\textit{#1}} % {\textrm{#1}}
\newcommand{\etal}{\textit{et~al}\@ifnextchar{\relax}{.\relax}{\ifx\@let@token.\else\ifx\@let@token~.\else.\@\xspace\fi\fi}}
\newcommand{\etc}{\foreign{etc}\@ifnextchar{\relax}{.\relax}{\ifx\@let@token.\else\ifx\@let@token~.\else.\@\xspace\fi\fi}}
\newcommand{\eg}{\foreign{e.g}\@ifnextchar{\relax}{.\relax}{\ifx\@let@token.\else\ifx\@let@token~.\else.\@\xspace\fi\fi}}
\newcommand{\ie}{\foreign{i.e}\@ifnextchar{\relax}{.\relax}{\ifx\@let@token.\else\ifx\@let@token~.\else.\@\xspace\fi\fi}}
\newcommand{\perse}{\foreign{per~se}\xspace}
\makeatother

\newcommand{\pN}{\text{PN}\xspace}

\definecolor{NoteColor}{rgb}{0.900, 0.218, 0.000}

\definecolor{NewColor}{rgb}{0,.55,0}

\newcommand{\software}[1]{\textsc{#1}}

\makeatletter
\newcommand{\CapName}[1]{\textbf{#1}.}
\newcommand{\ShowDimensions}{%
  \typeout{The font encoding is \f@encoding}        %
  \typeout{The font family is \f@family}            %
  \typeout{The font series is \f@series}            %
  \typeout{The font shape is \f@shape}              %
  \typeout{The font size is \f@size}                %
  \typeout{The baselineskip is \f@baselineskip}     %
  \typeout{The math font size is \tf@size}          %
  \typeout{The math script size is \sf@size}        %
  \typeout{The math scriptscript size is \ssf@size} %
  \typeout{The linewidth is \the\linewidth}         %
}
\newcommand{\prefixscripts}[2]{%
  \@mathmeasure\z@\displaystyle{#2}%
  \global\setbox\@ne\vbox to\ht\z@{}\dp\@ne\dp\z@
  \setbox\tw@\box\@ne
  \@mathmeasure4\displaystyle{\copy\tw@#1}%
  \@mathmeasure6\displaystyle{#2}%
  \dimen@-\wd6 \advance\dimen@\wd4 \advance\dimen@\wd\z@
  \hbox to\dimen@{}{\kern-\dimen@\box4\box6}%
}
\newcommand{\scripts}[3]{%
  \@mathmeasure\z@\displaystyle{#2}%
  \global\setbox\@ne\vbox to\ht\z@{}\dp\@ne\dp\z@
  \setbox\tw@\box\@ne
  \@mathmeasure4\displaystyle{\copy\tw@#1}%
  \@mathmeasure6\displaystyle{#2#3}%
  \dimen@-\wd6 \advance\dimen@\wd4 \advance\dimen@\wd\z@
  \hbox to\dimen@{}{\kern-\dimen@\box4\box6}%
}
\makeatother

\makeatletter
\let\protect\relax
{\catcode`\|=\active
  \xdef\InnerProduct{\protect\expandafter\noexpand\csname InnerProduct \endcsname}
  \expandafter\gdef\csname InnerProduct \endcsname#1{%
    \begingroup
    \ifx\SavedDoubleVert\relax
    \let\SavedDoubleVert\|\let\|\IpDoubleVert
    \fi
    \mathcode`\|32768\let|\IPVert
    \left({#1}\right)
    \endgroup
  }
}
\def\IPVert{\@ifnextchar|{\|\@gobble}% turn || into \|
     {\egroup\,\mid@vertical\,\bgroup}}
\def\IPDoubleVert{\egroup\,\mid@dblvertical\,\bgroup}
\let\SavedDoubleVert\relax
\def\midvert{\egroup\mid\bgroup}
\def\SetVert{\@ifnextchar|{\|\@gobble}% turn || into \|
    {\egroup\;\mid@vertical\;\bgroup}}
\def\SetDoubleVert{\egroup\;\mid@dblvertical\;\bgroup}
\def\mid@vertical{\mskip1mu\vrule\mskip1mu}
\def\mid@dblvertical{\mskip1mu\vrule\mskip2.5mu\vrule\mskip1mu}
\makeatother

\usepackage{xparse} % For \NewDocumentCommand, etc.

\makeatletter

\newcommand{\grading}[1]{\ensuremath{\langle #1 \rangle}}
\newcommand{\graded}[2]{\ensuremath{#1_{\grading{#2}}}}

\newcommand{\conjug@@te}[3]{\raisebox{#1}{$#2\text{\tiny{#3}}$}}
\newcommand{\conjug@te}[3]{\overline{#1}^{\mathpalette{\conjug@@te{#3}}{#2}}}
\newcommand{\reverse}[1]{\conjug@te{\multivec{#1}}{rev}{.4pt}}
\newcommand{\involution}[1]{\conjug@te{\multivec{#1}}{invo}{.4pt}}
\newcommand{\Cc}[1]{\conjug@te{\multivec{#1}}{Cc}{.1pt}}
\newcommand{\Herm}[1]{\conjug@te{\multivec{#1}}{Herm}{.1pt}}
\newcommand{\spinconjugate}[1]{\conjug@te{\multivec{#1}}{sc}{.4pt}}

\newcommand{\@multivector}[1]{\ensuremath{#1}}
\ExplSyntaxOn
\NewDocumentCommand{\multivec}{ mo }
    {
                \IfNoValueTF {#2}  { \@multivector{#1} }
                    { \graded{\@multivector{#1}}{#2} }
    }
\NewDocumentCommand{\multivecrev}{ mo }
    {
                \IfNoValueTF {#2}  { \reverse{\@multivector{#1}} }
                    { \graded{\reverse{\@multivector{#1}}}{#2} }
    }
\ExplSyntaxOff

\newcommand{\threevec}[1]{\ensuremath{\multivec{#1}}}
\newcommand{\unitvec}[1]{\ensuremath{\hat{#1}}}

\newcommand{\rotor}[1]{\ensuremath{\multivec{#1}}}

\DeclareMathAlphabet{\mathsfsl}{\encodingdefault}{\sfdefault}{m}{sl}

\definecolor{lightgrey}{gray}{0.75}
\def\maketag@@@#1{\@Restriction\hbox{\m@th\normalfont#1}}
\def\Restriction#1{\def\@Restriction{\llap{\textcolor{lightgrey}{$#1$}~}}}
\Restriction{}

\makeatother

\renewcommand{\threevec}[1]{\ensuremath{\vec{#1}}}

\newcommand{\wavefield}{h}

\newcommand{\AlignmentAxis}{\ensuremath{\hat{V}_{h}}}

\newcommand{\FieldAngularVelocity}{\ensuremath{\threevec{\omega}}}
\newcommand{\FieldAngularVelocityHat}{\ensuremath{\unitvec{\omega}}}
\newcommand{\LdtQuantity}{\ensuremath{\braket{L\, \partial_{t}}}}

\newcommand{\AngMomOp}{\ensuremath{L}}

\newcommand{\LLQuantity}{\ensuremath{\braket{LL}}}

\newcommand{\ParityInversion}{\ensuremath{P}}
\newcommand{\ParityInvariantProjection}{\ensuremath{\Pi}}
\newcommand{\ParityVariantProjection}{\ensuremath{\mathord{\amalg}}}
\newcommand{\AntipodalEvaluation}{\ensuremath{A}}
\newcommand{\ConjugateAntipodalEvaluation}{\ensuremath{\bar{\AntipodalEvaluation}}}
\newcommand{\ConjugateAntipodalInvariantProjection}{\ensuremath{\ParityInvariantProjection_{\ConjugateAntipodalEvaluation}}}

\newcommand{\NormalizedAntisymmetry}{\ensuremath{a}}
\newcommand{\NormalizedParityViolation}{\ensuremath{p_{\mathord{\min}}}}

\newcommand{\LuminosityDistance}{\ensuremath{R}} %d_{\text{L}}}}

\newcommand{\nhat}{\ensuremath{\unitvec{n}}}
\newcommand{\lambdahat}{\ensuremath{\unitvec{\lambda}}}
\newcommand{\ellhat}{\ensuremath{\unitvec{\ell}}}
\newcommand{\rhat}{\ensuremath{\unitvec{r}}}
\newcommand{\nHat}{\nhat}
\newcommand{\lambdaHat}{\lambdahat}
\newcommand{\ellHat}{\ellhat}
\newcommand{\rHat}{\rhat}

\newcommand{\xhat}{\ensuremath{\unitvec{x}}}
\newcommand{\yhat}{\ensuremath{\unitvec{y}}}
\newcommand{\zhat}{\ensuremath{\unitvec{z}}}
\newcommand{\xHat}{\xhat}
\newcommand{\yHat}{\yhat}
\newcommand{\zHat}{\zhat}
\newcommand{\Xhat}{\ensuremath{\unitvec{X}}}
\newcommand{\Yhat}{\ensuremath{\unitvec{Y}}}
\newcommand{\Zhat}{\ensuremath{\unitvec{Z}}}
\newcommand{\XHat}{\Xhat}
\newcommand{\YHat}{\Yhat}
\newcommand{\ZHat}{\Zhat}

\newcommand{\OmegaTot}{\ensuremath{\threevec{\Omega}_{\text{tot}}}}
\newcommand{\OmegaOrb}{\ensuremath{\threevec{\Omega}_{\text{orb}}}}
\newcommand{\OmegaTotHat}{\ensuremath{\unitvec{\Omega}_{\text{tot}}}}

\newcommand{\OmegaPrec}{\ensuremath{\threevec{\Omega}_{\text{prec}}}}

\newcommand{\chivec}[1]{\ensuremath{\threevec{\chi}_{#1}}}

\newcommand{\Rf}[1][]{\ensuremath{\rotor{R}_{\text{f#1}}}}

\newcommand{\Cornell}{\affiliation{Center for Radiophysics and
    Space Research, Cornell University, Ithaca, New York 14853, USA}} %

\newcommand{\CITA}{\affiliation{Canadian Institute for Theoretical
    Astrophysics, University of Toronto, 60 Saint George Street,
    Toronto, Ontario M5S 3H8, Canada}} %

\newcommand{\CIFAR}{\affiliation{Canadian Institute for Advanced
    Research, 180 Dundas St.~West, Toronto, ON M5G 1Z8, Canada}} %

\newcommand{\DAA}{\affiliation{Department of Astronomy and
    Astrophysics, 50 St.\ George Street, University of Toronto,
    Toronto, ON M5S 3H4, Canada}}

\begin{document}

%%%%%%%%%%%%%%%%

\graphicspath{%
  {Plots/}%
  % More directories are added in braces, without commas between
}

\title[Gravitational-wave modes from precessing binaries]
{Gravitational-wave modes from precessing black-hole binaries}

\makeatletter
\@booleantrue\frontmatterverbose@sw
\makeatother

\surnames{Boyle, Kidder, Ossokine, Pfeiffer}
\author{Michael Boyle} \Cornell %
\author{Lawrence E. Kidder} \Cornell %
\author{Serguei Ossokine} \CITA\DAA %
\author{Harald P. Pfeiffer} \CITA\CIFAR %

\newcommand{\arXivIdentifier}{1409.4431}
\newcommand{\ancillaryURL}{http://arxiv.org/src/\arXivIdentifier/anc/}
\newcommand{\paperNotebook}{GWModesFromPrecessingBHBinaries.ipynb}
\newcommand{\notebookViewerURL}{http://nbviewer.ipython.org/url/arxiv.org/src/\arXivIdentifier/anc/\paperNotebook}

\date{\today}

\begin{abstract}
  %%% Note: ArXiv limits the abstract to 1920 characters.  Some
  %%% macros have to be removed, and whitespace can be eliminated when
  %%% entering the abstract in the web form.  But this still leaves
  %%% the current abstract very close to the limit.  The accompanying
  %%% `abstract.txt` contains the shorted version of this abstract,
  %%% with all newlines and unnecessary macros removed.
  %
  Gravitational waves from precessing black-hole binaries exhibit
  features that are absent in nonprecessing systems.  The most
  prominent of these is a parity-violating asymmetry that beams energy
  and linear momentum preferentially along or opposite to the orbital
  angular momentum, leading to recoil of the binary.  The asymmetry
  will appear as amplitude and phase modulations at the orbital
  frequency.  For strongly precessing systems, it accounts for at
  least \SI{3}{\percent} amplitude modulation for binaries in the
  sensitivity band of ground-based gravitational-wave detectors, and
  can exceed \SI{50}{\percent} for massive systems.
  Such asymmetric features are also clearly visible when the waves are
  decomposed into modes of spin-weighted spherical harmonics, and are
  inherent in the waves themselves---rather than resulting from
  residual eccentricity in numerical simulations or from mode-mixing
  due to precession.  In particular, there is generically no
  instantaneous frame for which the mode decomposition will have any
  symmetry.
  We introduce a method to simplify the expressions for waveforms
  given in analytical relativity, which can be used to combine
  existing high-order waveforms for nonprecessing systems with
  expressions for the precessing contributions, leading to improved
  accuracy and a unified treatment of precessing and nonprecessing
  binaries.
  Using this method, it is possible to clarify the nature and the
  origins of the asymmetries and show the effects of asymmetry on
  recoils more clearly.  We present post-Newtonian (\pN) expressions
  for the waveform modes that include these terms, complete to the
  relative 2\pN level in spin (proportional to $v^{4}/c^{4}$ times a
  certain combination of the spins).  Comparing the results of those
  expressions to numerical results, we find good qualitative
  agreement.  We also demonstrate how these expressions can be used to
  efficiently calculate waveforms for gravitational-wave astronomy.
\end{abstract}


% 04.25.-g, % Approximation methods; equations of motion
% 04.25.D-, % Numerical relativity
% 04.25.dc, % NR studies of crit. behavior, sing.'s, cosmic censorsh.
% 04.25.dg, % NR studies of black holes and black-hole binaries
% 04.25.dk, % NR studies of other relativistic binaries
% 04.25.Nx, % PN approximation; perturbation theory; etc.
% 04.30.-w, % Gravitational waves
% 04.30.Db, % Wave generation and sources
% 04.30.Nk, % Wave propagation and interactions
% 04.30.Tv, % Gravitational-wave astrophysics
% 04.80.Nn, % Gravitational wave detectors and experiments

\maketitle

%%%%%%%%%%%%%%%%%%%%%%%%%%%%%%%%%%%%%%%%%%%%%%%%%%%%%%%%%%%%%%%%%%%%%%
%%%%%%%%%%%%%%%%%%%%%%%%%%%%%%%%%%%%%%%%%%%%%%%%%%%%%%%%%%%%%%%%%%%%%%
\section{Introduction}
\label{sec:Introduction}

The era of advanced gravitational-wave detectors will most likely
bring dozens to hundreds of detections of black-hole binaries per
year~\cite{LVC:2010}.  This raises the prospect of true
gravitational-wave astronomy, with which we will be able to explore
otherwise obscure regions of our universe.  But the power of such
exploration is limited by our ability to model the expected signals.
Unless we can accurately model the gravitational waves emitted by a
known or potential astrophysical source, we cannot know how sensitive
our detection pipeline is to that type of source.  Without
understanding how physical characteristics are imprinted onto a
waveform, we cannot expect to accurately measure any such
characteristic or even know the accuracy of an attempted
measurement---which diminishes the value of these detections to
science.  While simple approximate waveforms may be sufficient for
initial detections and explorations of data-analysis techniques,
gravitational-wave astronomy will require accurate
waveforms~\cite{raymondetal:2009, pekowskyetal:2013a,
  pekowskyetal:2013b}.

Precessing black-hole binaries form a particularly interesting class
of sources.  These are systems in which one or both black holes have
spin misaligned with the orbital axis, causing motion of that axis as
the binary evolves.  Precession encodes a wealth of information in the
gravitational-wave signal, which can break degeneracies and allow the
unambiguous measurement of various features of the
source~\cite{LVC:2013}.  Though the uncertainties are great, these
systems likely constitute a significant portion of potential
black-hole binaries to which advanced detectors will be
sensitive~\cite{ryan:1995, spruitphinney:1998, laietal:2001,
  buliketal:2004, grandclementetal:2004, buliketal:2004,
  wangetal:2007}.  Unfortunately, the richness of these signals
entails added complexity in the corresponding models.

\begin{figure*}
  % \tikzset{external/remake next=true}
  \includegraphics{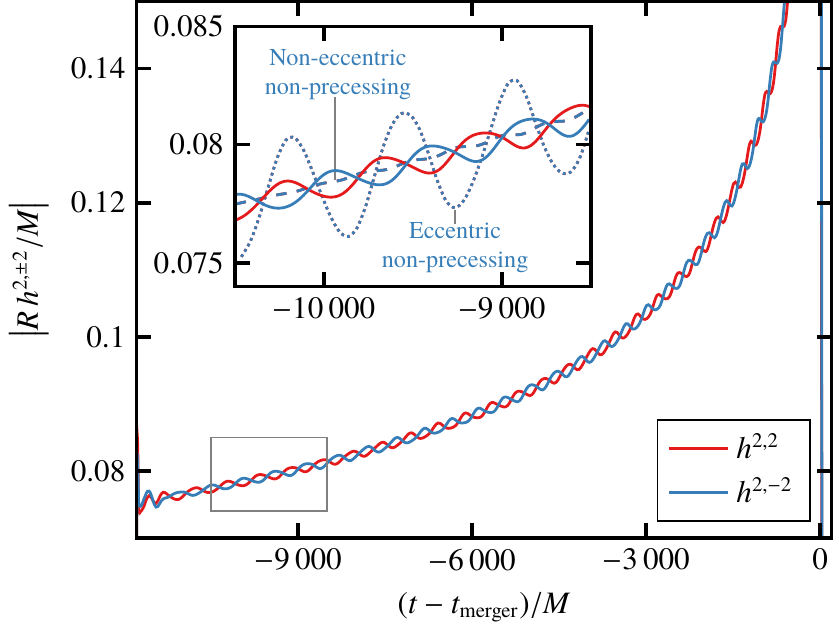} %
  \hfill
  % \tikzset{external/remake next=true}
  \includegraphics{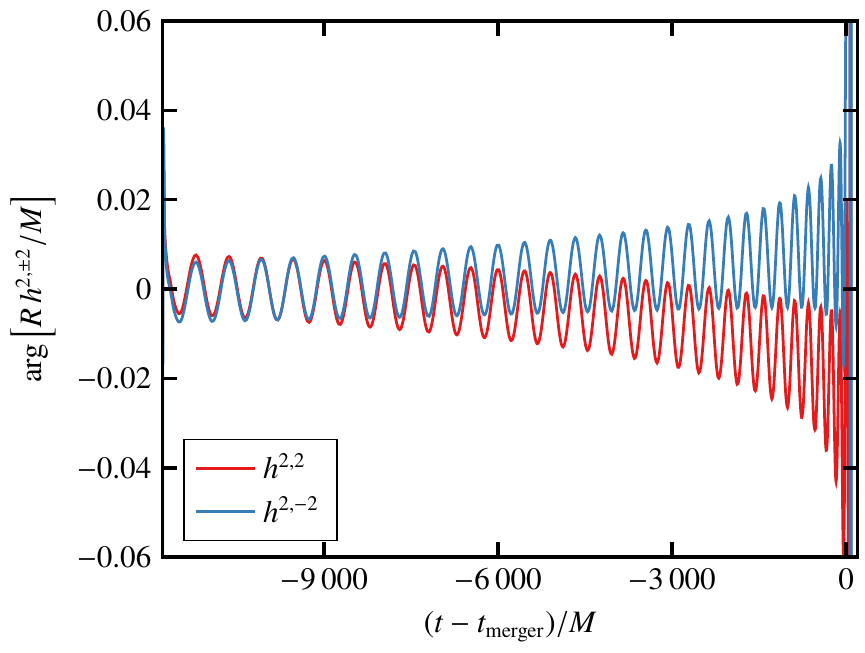} %
  \caption{ \label{fig:ModeAsymmetries} %
    \CapName{Dominant modes of a precessing system} %
    These plots show the amplitudes (left) and phases (right) of the
    $h^{2, \pm 2}$ modes from a numerical evolution of an equal-mass
    binary black-hole system with an initial spin of $\chi=0.5$ in the
    orbital plane on one hole, and no spin on the other.  For
    comparison, the inset of the left panel also show the curves for
    equal-mass nonspinning systems with very low eccentricity (dashed
    curve) and eccentricity $e \approx 0.02$ (dotted curve); the
    $h^{2,2}$ modes are plotted, but are identical to the $h^{2,-2}$
    modes for these nonprecessing systems.  %
    In all cases, the data are extrapolated to infinite
    radius~\cite{boylemroue:2009, tayloretal:2013} and are measured in
    the co-rotating frame~\cite{boyle:2013}, meaning that as much time
    dependence as possible has been removed from the waveform by a
    rotation of the coordinate system.  %
    The amplitude plot shows that the two modes of the precessing
    system alternate in dominance: when the $(\ell,m) = (2,2)$ mode
    dominates, more power is beamed above the orbital plane (along the
    angular velocity); when the $(2,-2)$ mode dominates, more power is
    beamed below the orbital plane (opposite to the angular velocity).
    The period of this oscillation is very nearly equal to the orbital
    period.  Similar features can be seen in other modes.  Notably,
    the period of oscillation is roughly independent of the $m$ value;
    it is always close to the orbital period.  %
    The phase plot exhibits oscillations on the same timescale.
    Because these oscillations are in phase with each other, it is
    clear that they could not be simultaneously eliminated from both
    modes---at least with a rotation about the $Z$ axis of the
    decomposition frame.  For example, if we were to transform to a
    frame in which the phase of the $h^{2,2}$ mode were constant, the
    oscillations in the phase of the $h^{2,-2}$ mode would be roughly
    twice as large.  The nonprecessing systems shown in the inset of
    the plot on the left would appear on this plot as very nearly
    constant curves close to $0$.  %
  }
\end{figure*}

Nonprecessing systems exhibit various symmetries, which reduce the
complexity of the systems.  For example, the black-hole spins are
essentially constant,\footnote{The spin directions should be precisely
  constant, while the spin magnitudes will experience a gradual (2\pN)
  change~\cite{alvi:2001}.} and two of the three orbital rotational
degrees of freedom are eliminated.  Moreover, essentially all
quantities will be smoothly monotonic and will vary on the inspiral
timescale.  Precessing systems, on the other hand, break all
symmetries.  The spins rotate, and all three degrees of orbital
rotational freedom are engaged.  More importantly, essentially all
quantities vary on orbital timescales.  Even in their simplest forms,
the dominant components of the gravitational waves---which are usually
quite smooth, monotonic, and symmetric---oscillate asymmetrically, as
seen in Fig.~\ref{fig:ModeAsymmetries}.
%%%%%%%%%%%%%%%%%%%%%%%%%%%%%%%%%%%%%%%%%%%%%%%%%%%%%%%%%%%%%%%%%%%%%%
%%%%%%%%%%%%%%%%%%%%%%%%%%%%%%%%%%%%%%%%%%%%%%%%%%%%%%%%%%%%%%%%%%%%%%

While modern numerical codes can simulate precessing black-hole
binaries robustly, the results are meaningless accumulations of
numbers unless we can relate them to analytical
models~\cite{campanellietal:2009}.  But this task is made far more
challenging by the lack of symmetry.  Our purpose here is to explore
some of the features unique to precessing systems, and show that the
problem can be made tractable by representing the data in appropriate
ways.  We will introduce new ways of measuring waveforms and a new way
of expressing analytical models.  In so doing, we will find that
analytical models can be efficient both conceptually and
computationally, while accurately reproducing the key features of
precessing waveforms.

To begin, we combine the usual components of the transverse-traceless
projection of the metric perturbation\footnote{The methods and
  conclusions of this paper are essentially unchanged when considering
  the Newman--Penrose quantity $\Psi_{4}$ in place of $\h$.  For
  simplicity, we will only discuss $\h$ explicitly.} at time $t$ and
location $\threevec{r}$ relative to the binary, $\hp(t, \threevec{r})$
and $\hc(t, \threevec{r})$, into a single complex quantity $\h(t,
\threevec{r}) \defined \hp(t, \threevec{r}) - \i\, \hc(t,
\threevec{r})$.  At each instant of time, $\h$ is measured on a
coordinate sphere, and we abuse notation slightly by discussing $\h(t,
\rHat)$ while suppressing the radius of the sphere.\footnote{In fact,
  notation is frequently abused further by using $\h$ to represent the
  leading-order behavior of $\lvert \threevec{r} \rvert\, \h$ as the
  radius of the sphere approaches infinity.}  Finally, we decompose
the angular dependence as an expansion in spin-weighted spherical
harmonics (SWSHs)~\cite{newmanpenrose:1966, goldbergetal:1967,
  ninja2:2011} so that
\begin{equation}
  \label{eq:SWSHExpansion}
  \h(t, \rHat) = \sum_{\ell,m} h^{\ell,m}(t)\, \mTwoYlm{\ell,m}(\rHat).
\end{equation}
Thus, we generally discuss the modes $h^{\ell,m}(t)$, rather than the
function value in any particular direction.  This representation has
the advantage of transforming simply under rotations---a crucial
feature when dealing with precessing systems.  In particular, if
$\Rotated{h}^{\ell,m}$ are the modes measured in a second coordinate
system---for instance, one that is adapted to the instantaneous
orbital plane---then we have
\begin{equation}
  \label{eq:ModeRotation}
  \Rotated{h}^{\ell, m} = \sum_{m'} h^{\ell, m'}\, \D^{(\ell)}_{m',m}
  \left( \rotor{R}^{-1} \right),
\end{equation}
where $\D^{(\ell)}$ is the usual Wigner matrix and $\rotor{R}$ rotates
the first set of basis vectors into the second.  We show in
Appendix~\ref{sec:WaveformsForDataAnalysis} that the value of the
field can be efficiently calculated from the modes in the rotating
coordinate system, without first going through the numerically
expensive transformation of Eq.~\eqref{eq:ModeRotation}---which leads
us to suggest that this may also be a useful representation of the
waveform in data analysis for gravitational-wave detectors.

One of the more familiar expressions for the modes in nonprecessing
binaries expresses the invariance of the system under reflection
across the orbital plane (generally taken to coincide with the $x$-$y$
plane):\footnote{This equation is derived and discussed in greater
  detail in Appendix~\ref{sec:Parityandantipodes}.}
\begin{equation}
  \label{eq:ModeSymmetry}
  h^{\ell,m} = (-1)^{\ell}\, \bar{h}^{\ell,-m}.
\end{equation}
This relationship obviates the need to separately analyze modes with
negative values of $m$, for example.  In particular, this means that
the complex amplitudes of the two modes $(\ell, \pm m)$ are equal, and
their complex phases are opposite (up to an addition of $\pi$ for odd
$\ell$).  Equation~\eqref{eq:ModeSymmetry} and its related properties
are, of course, not true of nonprecessing binaries when the orbital
plane does not coincide with the $x$-$y$ plane of the coordinate
system used to decompose the SWSHs, and is generally not true for
precessing binaries in any coordinate system---as exemplified in
Fig.~\ref{fig:ModeAsymmetries}.  Similarly, a common expression for
the dependence of the modes on orbital phase $\Phi$ is\footnote{It
  must be noted that this expression gives the dependence of
  $h^{\ell,m}$ \emph{on} the \emph{orbital} phase for nonprecessing
  systems, but is sometimes incorrectly understood to also give the
  behavior \emph{of} the \emph{complex} phase of $h^{\ell,m}$.  That
  is wrong even for nonprecessing systems at the 2.5\pN level, because
  the proportionality is given by a time-dependent complex factor.}
\begin{equation}
  \label{eq:OrbitalPhaseDependence}
  h^{\ell,m} \propto e^{-\i\, m\, \Phi}.
\end{equation}
Again, this is incorrect even for nonprecessing binaries if the
orbital plane and the $x$-$y$ plane do not coincide, and generally not
true for precessing binaries in any coordinate system.

The crucial fact in the failures of Eqs.~\eqref{eq:ModeSymmetry}
and~\eqref{eq:OrbitalPhaseDependence} for precessing binaries,
however, is that those equations are not incorrect just because of
mode mixing as the orbital plane precesses while the decomposition
basis is left fixed.  Even if we allow rotations of the decomposition
frame used to measure the waveform modes, we will see that there is no
frame in which these equations would be true.  Instead, even in the
simplest frame, both Eqs.~\eqref{eq:ModeSymmetry}
and~\eqref{eq:OrbitalPhaseDependence} are wrong for precessing systems
at the 1\pN level (proportional to $v^{2}$ times a certain combination
of the spins).  Since $v \gtrsim 0.1$ for essentially all black-hole
binary systems expected to be visible to advanced gravitational-wave
detectors, these discrepancies can have relative magnitudes of $v^{2}
\gtrsim \SI{1}{\percent}$, depending on the spin, even at the lowest
frequencies to which the detectors are sensitive---and they constitute
ever-increasing proportions of the signal as the system inspirals.  We
will also find terms contributing to $h^{\ell,m}$ for precessing
systems that are proportional to $e^{-i\, (m\pm1)\, \Phi}$; because of
these factors, the 1\pN amplitude effects will oscillate on orbital
timescales.

Several techniques have been introduced to simplify waveform modes by
rotating the frame with respect to which the modes are decomposed.
Whereas the waveforms may originally be decomposed with respect to a
static basis $(\xHat, \yHat, \zHat)$, a new frame is defined by
constructing another basis $(\XHat, \YHat, \ZHat)$ at each moment in
time, and expressing the modes $h^{\ell,m}$ with respect to this
dynamic basis.  We might distinguish two such types of frame
determined by the waveforms themselves.  First is the co-precessing
frame~\cite{schmidtetal:2011, oshaughnessyetal:2011, boyleetal:2011},
in which the waveform is still rotating; the $\ZHat$ axis is aligned
with a certain feature in the waveform, but the rotation of that frame
is otherwise minimized.  Second is the co-rotating
frame~\cite{boyle:2013}, in which the waveform is not rotating at all.
Both of these frames can be determined from the waveforms alone.

There are also two closely analogous frames that can be useful for \pN
systems.  These are defined with respect to the binary's orbital
elements---the positions and velocities of the black holes---rather
than the waveforms.  First is a frame we might call the
``co-nutating'' frame (for reasons that will become clear in
Sec.~\ref{sec:Waveformattitude}), in which the $\ZHat$ axis is aligned
with the orbital angular velocity, but the rotation of that frame is
otherwise minimized.  This frame was introduced in
Ref.~\cite{buonannoetal:2003} with an eye toward simplifying the
analysis of gravitational-wave data from detectors.
%% N.B.: The wording of this previous sentence is a little delicate,
%% because it's not clear how successful their results were; don't
%% change it without discussing the change with the other authors.
Here, we are concerned exclusively with constructing simple, accurate
models of gravitational waves.  We therefore introduce a final frame:
the ``co-orbital'' frame, in which the binary itself is not moving at
all.  We discuss this further in
Sec.~\ref{sec:Waveformsinthecoorbitalframe}.

Decomposing waveform modes in any of these four frames will indeed
simplify certain features.  However, the key point is that while a
rotation can introduce parity-violating asymmetries where there would
otherwise be none, no rotation can eliminate asymmetries in waveforms
from precessing systems, as we will show in
Sec.~\ref{sec:Invariantcombinationsofmodes}.  Moreover, the
asymmetries fluctuate on an orbital timescale.  Thus, \emph{accurate
  precessing waveforms must always have features varying on the
  orbital timescale,} regardless of the decomposition frame.

Fortunately, these features arise from terms that already appear in
the \pN literature~\cite{kidder:1995, schnittmanetal:2008,
  racineetal:2008, arunetal:2009, buonannoetal:2012, boheetal:2013},
though in somewhat obscure form and with little direct discussion of
their effects.  In this paper, we will use recent advances in the
treatment of waveforms from precessing systems to discuss these
features in detail, clarify their origins, and correct some
misconceptions that seem to have arisen in the literature.  We exhibit
the relevant \pN expressions for these effects in terms of the
$h^{\ell,m}$ modes using a simple and unified framework, at the
highest order currently available, so that they may be easily
incorporated into future work requiring accurate waveform models.

We begin in Sec.~\ref{sec:AsymmetriesInNumericalData} by simply
demonstrating several manifestations of the asymmetries in numerical
data---first, in a familiar but potentially ambiguous way; then, in
various geometrically unambiguous ways.  In particular, we introduce
rotationally invariant measures of asymmetry and parity violation.  In
Sec.~\ref{sec:Asymmetriesinpnresults}, we introduce the co-orbital
frame more precisely.  We use the co-orbital frame to express the \pN
waveforms for precessing systems, and to understand the origin of the
asymmetric features.  We then demonstrate that the same features seen
in the numerical data are also present in post-Newtonian waveforms
when these terms are included.  The impact on binary recoil is briefly
discussed in Sec.~\ref{sec:Recoil}, where the effects of parity
violation on the recoil are analyzed in detail.  Finally, we summarize
the discussion in Sec.~\ref{sec:Conclusions}.  Three appendices are
also included.  The first gives explicit formulas for the
contributions to the \pN waveform from terms involving spin, which
allow for immediate implementation.  The second appendix exhibits an
efficient method for evaluating the waveform an inertial observer
would measure, given a waveform in a rotating frame---by means of
which we can avoid $\roughly \num{1000}$ evaluations of elements of
the Wigner $\D^{(\ell)}$ matrices at each time step, while improving
the accuracy of the result.  The final appendix discusses various
details of antisymmetry and parity violation necessary for deriving
results used in Sec.~\ref{sec:Invariantcombinationsofmodes}.  Explicit
implementations of all the methods and results of this paper are also
provided as computer code included among the \href{\ancillaryURL}
{ancillary files} on this paper's arXiv page.

%%%%%%%%%%%%%%%%%%%%%%%%%%%%%%%%%%%%%%%%%%%%%%%%%%%%%%%%%%%%%%%%%%%%%%
%%%%%%%%%%%%%%%%%%%%%%%%%%%%%%%%%%%%%%%%%%%%%%%%%%%%%%%%%%%%%%%%%%%%%%
\section{Asymmetries in numerical data}
\label{sec:AsymmetriesInNumericalData}
We begin our discussion of the asymmetries by exhibiting them in data
from a numerical evolution of a representative precessing binary;
comparable features are found in other precessing systems.  We choose
a system in which one black hole is nonspinning and the other has a
dimensionless spin of $\chi=0.5$, initially in the orbital
plane.\footnote{ Specifically, this is run
  \href{http://www.black-holes.org/waveforms/data/DisplayMetadataFile.php/?id=SXS:BBH:0003}%
  {\texttt{SXS:BBH:0003}} described in Ref.~\cite{mroueetal:2013}.}
The masses are nearly equal, with a relative mass difference
$(M_{1}-M_{2})/(M_{1}+M_{2}) \approx 0.02$, where $M_1$ and $M_2$ are
the component masses.  The orbital eccentricity of this numerical
simulation is estimated to be $e\approx 3\times 10^{-4}$.
Oscillations in phase and amplitude due to orbital eccentricity for
this system are proportional to $e$~\cite{wahlquist:1987}, and are
much smaller for this system than the features visible in
Fig.~\ref{fig:ModeAsymmetries}.

We demonstrate several different aspects of the asymmetry.  First, we
discuss asymmetries in the context of the waveform as decomposed into
spin-weighted spherical-harmonic (SWSH) modes.  Though this is perhaps
the most familiar representation of gravitational waveforms, there may
be some concern, in that individual modes are not rotationally
invariant, and therefore do not provide a robust measure of asymmetry.
We therefore introduce rotationally invariant integrals of the
waveform---expressed as combinations of the modes---that unambiguously
describe the asymmetries.  Finally, we will examine various quantities
describing the geometry and dynamics of the waveform and of the binary
itself.

\subsection{Waveform modes}
\label{sec:Waveformmodes}
Figure~\ref{fig:ModeAsymmetries} shows the amplitudes (left panel) and
phases (right panel) of the $h^{2,\pm2}$ modes of the waveform.  These
quantities are measured in the co-rotating frame~\cite{boyle:2013}, in
which as much of the time dependence as possible is absorbed into a
time-dependent rotation.  Both plots show oscillations on the orbital
timescale, and substantial differences between the $h^{2,2}$ and
$h^{2,-2}$ modes.  A nonprecessing system would have very smooth
curves with no apparent features on the orbital timescale and the
amplitudes of the two modes would be identical.  Furthermore, for such
a nonprecessing binary represented in the co-rotating frame, the usual
increases of the phase by multiples of $2\pi$ per orbit are absent.
Any remaining variations of the phases would appear oppositely in the
$h^{2,2}$ and $h^{2,-2}$ modes, in accordance with
Eq.~\eqref{eq:ModeSymmetry} but in contrast to what is observed in
Fig.~\ref{fig:ModeAsymmetries}.  However, the features we see in the
precessing system are \emph{not} artifacts of the attitude; we will
see in Sec.~\ref{sec:Invariantcombinationsofmodes} that they cannot be
eliminated through rotation of the decomposition basis.

Because of the structure of the SWSHs, the $\mTwoYlm{\ell,m}$
with positive $m$ values have greater amplitude in
directions with positive $z$ values; harmonics with negative $m$
values have greater amplitude in directions with negative $z$ values.
(This is in contrast with the more familiar scalar spherical
harmonics, and is required for compatibility between the behavior of
spin-weighted functions and the naive tangent basis defined with
respect to spherical coordinates.)  As a result, whenever $\lvert
h^{2,2} \rvert > \lvert h^{2,-2} \rvert$, net energy and linear
momentum are beamed \emph{along} the orbital angular velocity; when
$\lvert h^{2,2}\rvert < \lvert h^{2,-2} \rvert$, more the net momenta
are beamed \emph{opposite} the angular velocity.

As shown by the formulas in
Appendix~\ref{sec:Waveformmodeswithasymmetriccontributions}, when the
orbital plane coincides with the $x$-$y$ plane, the relative amplitude
difference between the $(2,\pm 2)$ modes is given at lowest order in
\pN theory as
\begin{equation}
 \label{eq:RelativeAmplitudeDifference}
 \frac{\abs{h^{2,2}} - \abs{h^{2,-2}}} {\abs{h^{2,2}} +
   \abs{h^{2,-2}}} \approx -v^{2}\, \frac{\threevec{\Sigma} \cdot
   \lambdaHat} {2M^{2}}.
\end{equation}
Here, $\threevec{\Sigma}/M = M_{2} \chivec2 - M_{1} \chivec1$, with
$\chivec1$ and $\chivec2$ being the dimensionless spins of the two
black holes; $M$ is the sum of the two component masses; $v$ is the
standard \pN-expansion parameter---roughly the relative speed of the
black holes; and $\lambdaHat$ is a unit vector in the orbital plane,
orthogonal to the black-hole separation vector.

Because $\Sigma/M^2$ can be of order unity, the asymmetry between
$h^{2,2}$ and $h^{2,-2}$ will be substantial in the late stages of a
binary black hole inspiral, where the velocity approaches $v \approx
1$.  Even at the \SI{10}{\hertz} low-frequency ``seismic wall'' of
advanced earthbound gravitational-wave detectors~\cite{harryetal:2010,
  shoemaker:2010}, for the very low total mass of \SI{10}{\SolarMass}
the relative amplitude difference given by
Eq.~\eqref{eq:RelativeAmplitudeDifference} is \SI{0.7}{\percent} in
strongly precessing systems.  Furthermore, the size of this effect
will only grow as the system approaches merger, exceeding
\SI{3}{\percent} at the frequencies to which advanced LIGO will be
most sensitive.  Higher-mass systems will exhibit correspondingly
larger oscillations at the same frequencies.  Generally, we can expect
any system to have relative asymmetries of as much as \SI{8}{\percent}
at the innermost stable circular orbit (ISCO)~\cite{favata:2011},
which is generally taken as the point at which \pN approximations
break down.  The numerical data for the system shown here reach
relative differences of \SI{14}{\percent} just after merger.
Extrapolating with the scaling from \pN theory, this suggests that
strongly precessing systems could exhibit differences greater than
\SI{50}{\percent}.

Though the $h^{2,\pm2}$ modes shown here exhibit the largest
oscillations and asymmetries in an absolute sense, higher harmonics
exhibit larger effects relative to their overall amplitudes.
Generally, we can conclude that precessing systems exhibit strong
amplitude and phase modulations throughout their evolution.  These
features must be modeled if we wish to obtain accurate waveforms and
extract accurate physics.

\subsection{Rotationally invariant measures of asymmetry}
\label{sec:Invariantcombinationsofmodes}
The complicated transformation law of Eq.~\eqref{eq:ModeRotation}
suggests that we cannot expect the relative amplitude difference given
in Eq.~\eqref{eq:RelativeAmplitudeDifference} to be rotationally
invariant.  For example, we could flip the sign of the left-hand side
of Eq.~\eqref{eq:RelativeAmplitudeDifference} by rotating $\zHat$ into
$-\zHat$.  It is natural to wonder if we could remove the asymmetries
entirely simply by rotating the system.  Here, we introduce two
rotationally invariant measures of the asymmetry; because of their
invariance and the fact that they are nonzero for precessing systems,
this demonstrates that no rotation can remove the asymmetry.

There are two important qualities of the asymmetry shown in
Fig.~\ref{fig:ModeAsymmetries}.  First, is the simple fact that the
magnitude of the waveform in one direction is different from the
magnitude in the opposite direction---its antipode.  We introduce the
antipodal operator $\AntipodalEvaluation$, which transforms a field
into that field evaluated at the antipodes.  Let $f(\rHat)$ be a
function defined on the unit sphere (\eg, a SWSH).  For any direction
$\rHat$, we define
\begin{equation}
  \label{eq:AntipodalOperator}
  \AntipodalEvaluation\{f\}(\rHat) \defined f(-\rHat).
\end{equation}
We will, of course, be most interested in fields of spin weight
$s=-2$.  As shown in Appendix~\ref{sec:Parityandantipodes},
$\AntipodalEvaluation$ reverses the spin weight of such fields.  To
ensure that our results behave properly under rotations, we must
reverse the spin weight again by taking the complex conjugate of the
field.  In particular, we define the \emph{conjugate} antipodal
operator
\begin{equation}
  \label{eq:ConjugateAntipodalOperator}
  \ConjugateAntipodalEvaluation\{f\}(\rHat) \defined \bar{f}(-\rHat).
\end{equation}
We can now apply this to the particular case of $f=h$.  We drop the
time dependence of $h$, stipulating that the following formulas apply
separately at each instant of time.  The effect of
$\ConjugateAntipodalEvaluation$ on the waveform modes is calculated in
Appendix~\ref{sec:Parityandantipodes}, giving the fairly simple
relation
\begin{equation}
  \label{eq:ConjugateAntipodalOperator_Modes}
  \ConjugateAntipodalEvaluation\{h\}^{\ell,m} = (-1)^{\ell+m}\,
  \bar{h}^{\ell,-m}.
\end{equation}
We define the projection operator
$\ConjugateAntipodalInvariantProjection \defined \frac{1}{2}(1 -
\ConjugateAntipodalEvaluation)$, which leaves only the antisymmetric
part of the waveform.  Using the involution property
$\ConjugateAntipodalEvaluation^{2} = 1$, it is trivial to compute that
$\ConjugateAntipodalInvariantProjection \{h\}$ is an eigenfunction of
$\ConjugateAntipodalEvaluation$ with eigenvalue $-1$; that is, we have
succeeded in extracting that part of the waveform that reverses sign
under $\ConjugateAntipodalEvaluation$.  We then use this projection to
define the normalized antisymmetry
\begin{subequations}
  \label{eq:NormalizedAsymmetry}
  \begin{align}
    \label{eq:NormalizedAsymmetryIntegral}
    \NormalizedAntisymmetry &\defined \sqrt{ \frac{\int
        \abs{\ConjugateAntipodalInvariantProjection\{h\}}^{2} \d
        \Omega} {\int \abs{h}^{2} \d \Omega} } \\
    \label{eq:NormalizedAsymmetryModes}
    &= \sqrt{ \frac{\sum_{\ell,m} \abs{h^{\ell,m} - (-1)^{\ell+m}
          \bar{h}^{\ell,-m}}^{2}} {4\sum_{\ell,m} \abs{h^{\ell,m}}^{2}}}.
  \end{align}
\end{subequations}
Though it is less familiar than the $h^{2,\pm2}$ modes seen above,
this quantity has the advantages of being rotationally
invariant\footnote{To understand the behavior of $h$ and $A\{h\}$
  under rotations, it is sufficient to consider the value of the
  fields measured at $\rHat$ and $-\rHat$ when the rotation is about
  that axis.  For spin-weighted fields, the rotation will induce
  opposite phase rotations, whereas we need the phases to vary in the
  \emph{same} way if their difference is to be independent of
  attitude.  This is why we need to use the complex conjugate in
  defining $\ConjugateAntipodalInvariantProjection$.  It is also easy
  to use the transformation law~\eqref{eq:ModeRotation} and various
  properties of the $\D^{(\ell)}$ matrices to show explicitly that the
  particular combination of modes seen in
  Eq.~\eqref{eq:NormalizedAsymmetryModes} is independent of the
  rotation operator $\rotor{R}$.  The proof is given in the
  \href{\notebookViewerURL\#II.-B.-Rotationally-invariant-measures-of-asymmetry}
  {ancillary files}.} and incorporating information about the complete
function, rather than just a few select modes.  This antisymmetry is
also related to the binary recoil due to emission of linear momentum
in the form of gravitational waves, as discussed further in
Sec.~\ref{sec:Recoil}.

The antisymmetry $\NormalizedAntisymmetry$ can be nonzero in
nonprecessing systems if the masses or spins of the black holes are
unequal~\cite{fitchett:1983, blanchetetal:2005, damourgopakumar:2006,
  schnittmanetal:2008}.  The second quality of asymmetry we wish to
discuss is one that is found only in precessing systems: inherent
parity violation.  In particular, a nonprecessing system will be
symmetric under reflection through the $x$-$y$ plane.  This
``$z$-parity'' operation is distinct from the closely related standard
parity operation in three-dimensional physics, which reverses the sign
of all spatial vector components; the two are related by an additional
rotation through $\pi$ about the $z$ axis.  Taking into account the
spin weight, Appendix~\ref{sec:Parityandantipodes} shows that the
effect of the $z$-parity operator $\ParityInversion_{z}$ on $h$ is
\begin{equation}
  \label{eq:ZParityOperator}
  \ParityInversion_{z}\{h\}(\rHat) = \bar{h}
  \left( \ParityInversion_{z} \left\{ \rHat \right\} \right).
\end{equation}
Again, we can find a simple expression in terms of the effect on the
waveform modes:
\begin{equation}
  \label{eq:ZParityOperator_Modes}
  \ParityInversion_{z}\{h\}^{\ell,m} = (-1)^{\ell}\, \bar{h}^{\ell,-m}.
\end{equation} %
[This shows that Eq.~\eqref{eq:ModeSymmetry} is just the statement
that $h = \ParityInversion_{z}\{h\}$ for nonprecessing systems when
the orbital plane is orthogonal to $\zHat$.]  And again, we construct
a projection operator $\ParityInvariantProjection_{z} \defined
\frac{1}{2}(1-\ParityInversion_{z})$, which leaves only the part of
the waveform that is antisymmetric under $\ParityInversion_{z}$.
Unfortunately, the result of this operator does not behave well under
rotations---note the crucial factor of $(-1)^{m}$ in
Eq.~\eqref{eq:ConjugateAntipodalOperator_Modes}.  This is a natural
consequence of the special choice of $z$ axis in the definition of
$\ParityInversion_{z}$, and presents an obstacle to defining a
rotationally invariant measure of parity violation analogous to the
antisymmetry $\NormalizedAntisymmetry$.  It is worth recalling that
our intention is to show that this parity violation is present in any
frame used to measure the waveform.  We can achieve that goal and
ensure rotational invariance by defining the normalized parity
violation as the \emph{minimum} such value, over all possible
attitudes $\mathcal{R}$:\footnote{We know of no way to eliminate the
  minimization process; as far as we know, some explicit numerical
  optimization is necessary.  However, the problem is essentially
  two-dimensional, rather than the naive expectation of
  three-dimensional, because the result is insensitive to a final
  rotation about the $z$ axis.  This makes the procedure far more
  efficient.  See the \href{\ancillaryURL} {ancillary materials} for
  more detail.}
\begin{subequations}
  \label{eq:NormalizedZParityViolation}
  \begin{align}
    \label{eq:NormalizedZParityViolationIntegral}
    \NormalizedParityViolation &\defined \min_{\mathcal{R}} \sqrt{
      \frac{\int \abs{\ParityInvariantProjection_{z}
          \big\{\mathcal{R}\{h\}\big\}}^{2} \d \Omega} {\int
        \abs{\mathcal{R}\{h\}}^{2} \d \Omega} } \\
    \label{eq:NormalizedZParityViolationModes}
    &= \min_{\mathcal{R}} \sqrt{ \frac{\sum_{\ell,m}
        \abs{\mathcal{R}\{h\}^{\ell,m} - (-1)^{\ell}
          \smash[t]{\overline{\mathcal{R}\{h\}}}^{\ell,-m}}^{2}}
      {4\sum_{\ell,m} \abs{h^{\ell,m}}^{2}}}.
  \end{align}
\end{subequations}
Because of this minimization, the result will be independent of the
original attitude of the frame used to measure $h$ by construction.
More specifically, when this value is nonzero---as in our chosen
system---we know that there cannot be any frame in which
Eq.~\eqref{eq:ModeSymmetry} is satisfied.

%%%%%%%%%%%%%%%%%%%%%%%%%%%%%%%%%%%%%%%%%%%%%%%%%%%%%%%%%%%%%%%%%%%%%%
\begin{figure}
  % \tikzset{external/remake next=true}
  \includegraphics{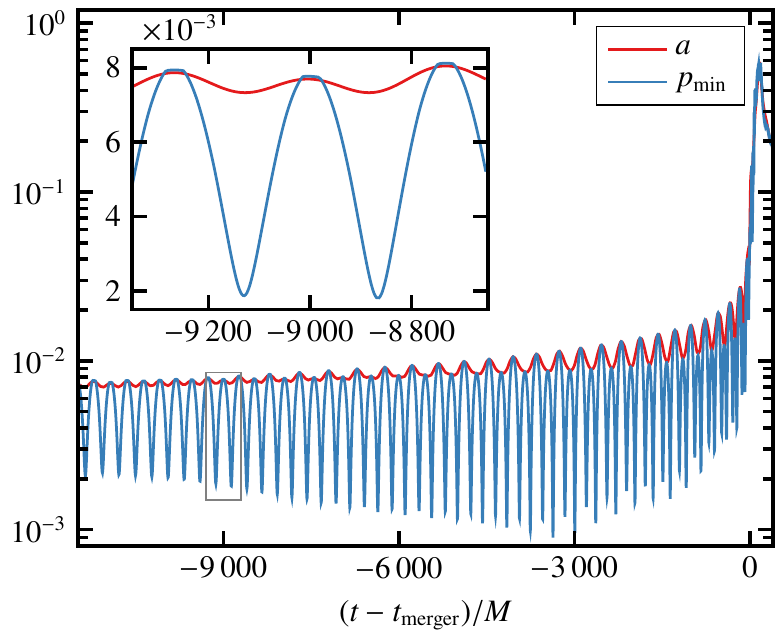} %
  \caption{ \label{fig:Asymmetry} %
    \CapName{Rotationally invariant measures of asymmetry} This plot
    shows the antisymmetry $\NormalizedAntisymmetry$ defined in
    Eq.~\eqref{eq:NormalizedAsymmetry} and the minimal parity
    violation $\NormalizedParityViolation$ defined in
    Eq.~\eqref{eq:NormalizedZParityViolation}, for the system
    described at the beginning of
    Sec.~\ref{sec:AsymmetriesInNumericalData}.  Because
    $\NormalizedParityViolation\neq 0$, there is no frame in which the
    $h^{\ell,m}$ modes of this system satisfy
    Eq.~\eqref{eq:ModeSymmetry}.  %
  }
\end{figure}
%%%%%%%%%%%%%%%%%%%%%%%%%%%%%%%%%%%%%%%%%%%%%%%%%%%%%%%%%%%%%%%%%%%%%%
Figure~\ref{fig:Asymmetry} shows the antisymmetry
$\NormalizedAntisymmetry$ and minimal parity violation
$\NormalizedParityViolation$ for our chosen numerical system.  Both
curves are almost entirely determined by the $h^{2,\pm2}$ and
$h^{2,0}$ modes, as measured in the co-rotating frame.  The
antisymmetry starts out at roughly \SI{0.7}{\percent} at the beginning
of the simulation, and increases very rapidly near merger, reaching
\SI{6}{\percent} at the moment of merger, peaking near
\SI{60}{\percent} shortly after merger.  We note that the system shown
here exhibits relatively modest precession; considerations from \pN
theory suggest that we may expect such antisymmetries to be roughly
four times larger during the inspiral of strongly precessing systems.

Intriguingly, the axis which minimizes $\NormalizedParityViolation$ in
Eq.~\eqref{eq:NormalizedZParityViolation} always lies extremely close
to one of the basis axes of the co-rotating frame; during the
inspiral, this optimal axis sometimes switches discontinuously to a
direction close to a different basis axis of the co-rotating frame.
We can see this in the inset of Fig.~\ref{fig:Asymmetry}, which
represents a little over one full orbit.  There are brief periods,
twice per orbit, during which $\NormalizedParityViolation$ and
$\NormalizedAntisymmetry$ nearly agree---in fact,
$\NormalizedParityViolation$ is slightly larger, presumably due to the
influence of modes with odd $m$.  These correspond to times during
which the $h^{2,\pm2}$ modes are nearly equal and the minimal
parity-violation axis is very nearly the $\ZHat$ axis of the
co-rotating frame.  There is then a discontinuous change in the slope
of the $\NormalizedParityViolation$ curve, as the parity violation
along $\ZHat$ remains large, but the violation along $\YHat$ drops, so
the minimization of Eq.~\eqref{eq:NormalizedZParityViolation} switches
to that axis.  The parity projection along $\YHat$ is insensitive to
the $h^{2,\pm2}$ asymmetry.

While $\NormalizedAntisymmetry$ and $\NormalizedParityViolation$ are
\emph{rotationally} invariant, they are not \emph{translationally}
invariant.  In fact, the numerical data shown here come from a
numerical simulation with a non-zero total velocity in the initial
data.  We have removed this initial velocity for all data shown in
this paper by transforming the asymptotic waveform data to counteract
the velocity~\cite{boyleetal:2014c}.\footnote{That velocity is set to
  the ADM momentum divided by the ADM energy, where those quantities
  are measured in the initial data~\cite{arnowittetal:2008,
    cookpfeiffer:2004, lovelaceetal:2008, lovelace:2009,
    garciaetal:2012}.  This is appropriate under the assumption that
  the initial-data slice contains no significant contribution to the
  ADM momentum and energy from anything other than the black holes
  themselves---for example junk radiation or gravitational waves
  intentionally included in the initial data.}  The residual velocity
of the initial data only has magnitude $\roughly 6 \times 10^{-5}\,c$,
and the boost transformation \perse does not change the data
appreciably.  However, the resulting translation does have a
significant effect on the waveform, reducing the asymmetry by an order
of magnitude late in the inspiral despite the fact that the
displacement is less than $\roughly 1\,M$ throughout the simulation.
And although we set the initial velocity to be roughly zero, a recoil
(hence also translation) develops in the data as the system approaches
merger, which has noticeable effects.  For example, in
Fig.~\ref{fig:Asymmetry}, successive peak values of
$\NormalizedParityViolation$ are roughly equal near the beginning of
the simulation; closer to merger, successive peaks are distinctly
uneven.  These are entirely consistent with the effects of
translation.

The boost and spatial translation \emph{cannot} be eliminated by
extrapolation to infinite radius~\cite{boylemroue:2009},
Cauchy-characteristic evolution~\cite{tayloretal:2013,
  reisswigetal:2012, winicour:2012, bishopetal:1997, gomezetal:2007},
or any similar scheme.  Rather, like time-translation and rotation,
they are asymptotic symmetries of asymptotically flat
spacetimes,\footnote{In fact, the translations are part of a larger
  class of symmetries, deemed ``supertranslations''~\cite{sachs:1962b,
    newmanpenrose:1966, geroch:1981, draystreubel:1984,
    moreschi:1986}.  Combined with rotations and boosts, these
  comprise the general asymptotic symmetries of asymptotically flat
  spacetime---referred to as the Bondi--Metzner--Sachs (BMS)
  group~\cite{bondietal:1962, sachs:1962b, newmanpenrose:1966}.} and
thus correspond to inherent gauge freedoms.  We have simply chosen to
impose a gauge condition on the NR data to coincide roughly with the
\pN gauge early in the simulation.

\subsection{Waveform attitude}
\label{sec:Waveformattitude}
In addition to the rotationally invariant scalars introduced in the
last section, we can also examine five quantities that transform as
vectors under rotation.  Three are defined with respect to the
gravitational waves themselves.  To define the first, we need the
matrix~\cite{oshaughnessyetal:2011}
\begin{equation}
  \label{eq:AngularMomentumMatrix}
  \LLQuantity^{a b} \defined \sum_{\ell, m, m'}\,
  \co{\wavefield}^{\ell,m'} \braket{\ell, m'| \AngMomOp^{(a}\,
    \AngMomOp^{b)} |\ell, m}\, \wavefield^{\ell, m}.
\end{equation}
Here, the $\ket{\ell, m}$ represent the spin-weight $s=-2$ SWSH, on
which the angular-momentum operator $L^a$ acts just as in the
non-spin-weighted case~\cite{goldbergetal:1967}.  Our first vector is
the dominant eigenvector of this matrix, labeled
$\AlignmentAxis$.\footnote{Because it is an eigenvector, the sign of
  $\AlignmentAxis$ is meaningless; we always choose it to lie more
  parallel than antiparallel to $\FieldAngularVelocity$, defined in
  Eq.~\eqref{eq:AngularVelocity}.}  This can be thought of as the
approximate symmetry axis of the waveform.  We can also define
another vector explicitly:
\begin{equation}
  \label{eq:AngularMomentumVector}
  \LdtQuantity^{a} \defined \sum_{\ell, m, m'}\, \Im \left[
    \co{\wavefield}^{\ell, m'} \braket{\ell, m'| \AngMomOp^{a}
      |\ell, m}\, \partial_{t} \wavefield^{\ell,m} \right].
\end{equation}
This quantity has the interpretation of the time derivative of the
waveform projected into the ``rotational'' parts of the
waveform~\cite{boyle:2013}, and is equal to the angular-momentum
flux~\cite{winicour:1980, campanellietal:1999, loustozlochower:2007,
  ruizetal:2008, loustozlochower:2014}.  Finally, it is easy to derive
the angular velocity of the waveform~\cite{boyle:2013} using these
expressions:\footnote{Incidentally, this is the angular velocity
  integrated to obtain the co-rotating frame used in plotting the
  modes in Fig.~\ref{fig:ModeAsymmetries}.}
\begin{equation}
  \label{eq:AngularVelocity}
  \FieldAngularVelocity = -\LLQuantity^{-1} \cdot \LdtQuantity.
\end{equation} %
%%%%%%%%%%%%%%%%%%%%%%%%%%%%%%%%%%%%%%%%%%%%%%%%%%%%%%%%%%%%%%%%%%%%%%
\begin{figure}
  % \tikzset{external/remake next=true}
  \includegraphics{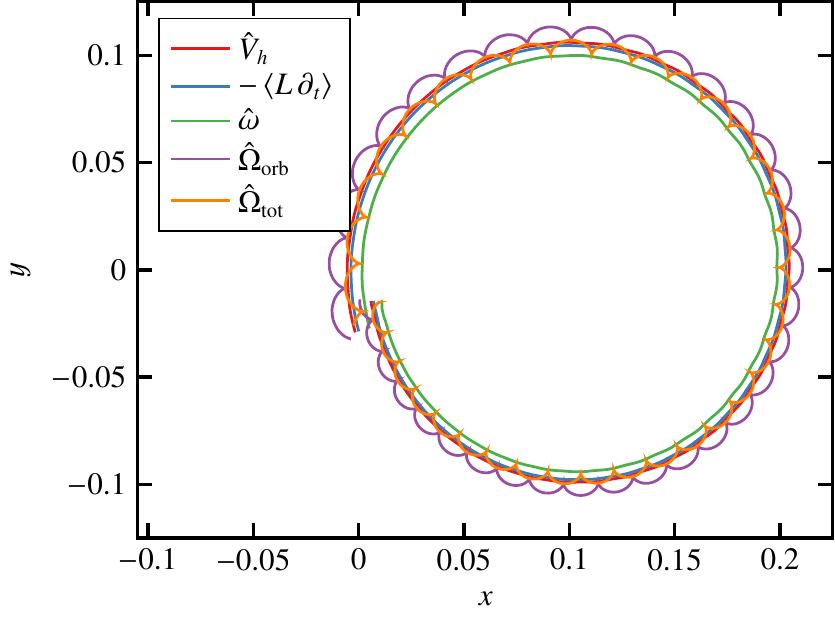} %
  \caption{ \label{fig:WaveformVectors} %
    \CapName{Vectors of a precessing system} This plot shows the
    components of five important vectors for a precessing system.  The
    curves evolve counterclockwise.
    The smooth curves are, from outer to inner, the waveform
    axis $\AlignmentAxis$, the projection of waveform time dependence
    onto rotation $\LdtQuantity$, and the waveform angular velocity
    $\FieldAngularVelocityHat$.
    The scalloped curves are derived from the numerical coordinates
    and show the nutations of the precession.  The outer curve is the
    \emph{orbital} angular velocity of the binary $\ellHat$; the inner
    one is the \emph{total} angular velocity $\OmegaTotHat$.
    The data shown are from the simulation described at the beginning
    of Sec.~\ref{sec:AsymmetriesInNumericalData}, but only showing the
    first precession cycle (roughly the first $8000\,M$) for clarity.
    In a nonprecessing system, all five vectors coincide; here they
    are clearly distinct.  We will see in
    Sec.~\ref{sec:Asymmetriesinpndata} that similar features are found
    in post-Newtonian results.  %
  }
\end{figure}
%%%%%%%%%%%%%%%%%%%%%%%%%%%%%%%%%%%%%%%%%%%%%%%%%%%%%%%%%%%%%%%%%%%%%%
The directions of these three vectors are plotted in
Fig.~\ref{fig:WaveformVectors}, appearing as the smooth curves.
Intriguingly, none of the three vectors are aligned with any other at
any time.  Our usual notion of the waveform rotating about its
alignment axis $\AlignmentAxis$ is incorrect.  And the
rotation axis of the waveform does not coincide with the axis of
angular-momentum flux.

For comparison, we also plot the directions of two vectors relating to
the dynamics of the binary, defined in terms of the coordinate
positions of the black holes.  Though these quantities are obviously
gauge dependent, we will find it useful to make contact with \pN
theory using these vectors; the results from the numerical simulation
are strikingly similar to the \pN results.  The orbital angular
velocity $\OmegaOrb$ is given by the usual expression
\begin{equation}
  \label{eq:OmegaOrb}
  \OmegaOrb=\unitvec{n} \times \dot{\unitvec{n}},
\end{equation}
where $\unitvec{n}$ is the separation vector between the two black
holes.  It is orthogonal to the orbital plane by definition, and
describes the instantaneous velocity of the binary.  We also need to
distinguish the \emph{total} angular velocity of the system
$\OmegaTot$.  For precession to occur, $\OmegaTot$ must also have a
component $\OmegaPrec$ along $\unitvec{n}$; essentially this
additional component gives the angular velocity of the orbital angular
velocity vector.  We have
\begin{equation}
  \label{eq:OmegaTot}
  \OmegaTot = \OmegaOrb + \OmegaPrec.
\end{equation}
The directions of these two vectors are also plotted in
Fig.~\ref{fig:WaveformVectors}, where they appear as the scalloped
curves, as a result of the nutations of the system.

It is interesting to note that the dynamics of the system somehow
conspire to eliminate the pronounced nutations of the orbital plane
from the gravitational radiation, resulting in relatively smooth
curves for all of the waveform vectors---though some oscillations are
still visible on close inspection.  This can be understood in terms of
the analysis of Ref.~\cite{grallaetal:2010}, which used simple
analogous systems to explain the nutations\footnote{The analysis of
  Ref.~\cite{grallaetal:2010} suggests that systems whose spin
  components in the plane are aligned should exhibit nutations, in
  which the orbital plane tilts on a time scale much faster than
  precession; with spins anti-aligned, the effect would look more like
  ``bobbing'', in which the orbital plane would move up and down.} as
occurring due to variations in the energy of different parts of a
spinning object in motion transverse to the spin vector.  The energy
of the portion of the body in prograde motion increases, while the
energy of the portion in retrograde motion decreases, so the center of
mass-energy shifts depending on the direction of motion relative to
the spin.  But since the spin is roughly constant on an orbital
timescale, this effect oscillates: the naive ``coordinate'' center of
the black hole (as measured with reference to the horizon) moves
relative to the center of mass-energy.  Evidently, the centers of
mass-energy are more relevant to the dynamics of the system, so we
expect these to move on relatively smoother trajectories, while the
coordinate centers nutate on the orbital timescale.  Indeed, the black
holes appear to nutate when considering only the coordinate positions
of the horizons, as seen in the scalloped curves of
Fig.~\ref{fig:WaveformVectors}.  It is also evidently the non-nutating
mass-energy that acts as the source of gravitational-wave emission,
which is why the three curves in Fig.~\ref{fig:WaveformVectors}
measured from the waveforms are smooth.

The most basic point to take away from Fig.~\ref{fig:WaveformVectors},
though, is that there is no simple relationship between the directions
of the orbital elements and various features in the waveforms.  We
will see in following sections that, to account for the non-alignment
between the various curves in the figure, we must retain the
asymmetric \pN mode terms responsible for the effects shown above.

%%%%%%%%%%%%%%%%%%%%%%%%%%%%%%%%%%%%%%%%%%%%%%%%%%%%%%%%%%%%%%%%%%%%%%
%%%%%%%%%%%%%%%%%%%%%%%%%%%%%%%%%%%%%%%%%%%%%%%%%%%%%%%%%%%%%%%%%%%%%%

\section{Asymmetries in \pN theory}
\label{sec:Asymmetriesinpnresults}

Each of the asymmetries demonstrated above can already be found in the
\pN literature, in some form, though they are often obscure.  To
improve this situation, we first present a simplified framework for
expressing the modes of the waveform, which allows for more tractable
analytical expressions, as well as a robust, accurate and unified
treatment of both precessing and nonprecessing systems.  We then use
this framework to describe the origins of the asymmetries in \pN
theory.  Finally, we compare the results of \pN calculations with the
numerical results shown above.

\subsection{Waveforms in the co-orbital frame}
\label{sec:Waveformsinthecoorbitalframe}

The standard framework for analyzing the radiation field of a
black-hole binary uses symmetric trace-free (STF)
tensors~\cite{thorne:1980}.  By taking various contractions between
these tensors and vectors describing the position and attitude of
an observer, we obtain the gravitational-wave field at the location of
that observer.  Alternatively, we can obtain the SWSH modes of the
field by contracting with certain other STF spherical-harmonic
tensors~\cite{thorne:1980, blanchet:2014}:
\begin{equation}
  \label{eq:WaveformModesAsTensors}
  h^{\ell,m} \propto \left( U_{L} + \i\frac{2\ell}{\ell+1}\, V_{L}
  \right)\, \mathcal{Y}^{\ell,-m}_{L}.
\end{equation}
Here, $U_{L}$ and $V_{L}$ are referred to as radiative mass- and
current-multipole tensors, the $\mathcal{Y}^{\ell,m}_{L}$ are
spherical-harmonic tensors, and there is an implied summation over
possible values of the multi-index $L$, which represents $\ell$ tensor
components.  $U_{L}$ and $V_{L}$ contain information about the physics
of the system, whereas $\mathcal{Y}^{\ell,m}_{L}$ describes how a
chosen coordinate system on the sphere relates to the modes.
Therefore, choosing a coordinate system that relates directly to the
physics can simplify the expressions for the modes; the modes can then
be rotated to any other frame using Eq.~\eqref{eq:ModeRotation}.

The rank-$\ell$ tensors $\mathcal{Y}^{\ell,m}_{L}$ are given
explicitly by Thorne~\cite{thorne:1980}.  Rather than reproducing the
general expression here, we simply describe the most important
features.  Each such tensor contains $m$ factors of $(\xHat - \i\,
\yHat)$ [or $-m$ factors of $(\xHat + \i\, \yHat)$ for $m<0$].  This
factor is multiplied by a sum of terms involving $n$ factors of the
form $\delta_{jk}$, for natural numbers $n \in \{0, \ldots, \lfloor
(\ell-m)/2 \rfloor\}$, with remaining factors given by $\zHat$ as
necessary for the tensor to have rank $\ell$.  There are two
conclusions we need to draw from this.  First is the obvious fact that
these tensors are given in terms of the $(\xHat, \yHat, \zHat)$ basis;
contractions with arbitrary vectors could lead to very complicated
expressions.  Second, $U_{L}$ or $V_{L}$ can only contribute to a
given $h^{\ell,m}$ mode if it has $\abs{m}+2n$ tensor components in
the $\xHat$-$\yHat$ plane for some natural number $n$.

The waveform multipole tensors $U_{L}$ and $V_{L}$ are generally given
in terms of a vector basis $(\nHat, \lambdaHat, \ellHat)$ and the spin
vectors of the black holes~\cite{thorne:1980, finnchernoff:1993,
  apostolatosetal:1994, willwiseman:1996, kidder:1995,
  buonannoetal:2003, buonannoetal:2012, blanchet:2014, boheetal:2013,
  blanchet:2014}.  Here, $\nHat$ is a vector pointing from one black
hole to the other; $\lambdaHat$ is parallel to the time derivative
$\dot{\nHat}$; and $\ellHat = \nHat \times \lambdaHat$ is parallel to
the orbital angular velocity $\OmegaOrb$.  For example, we have the
lowest-order quadrupole contributions\footnote{Angle brackets indicate
  the symmetric trace-free (STF) part of the tensor, but that is
  unimportant to the argument; the tensors with which these are
  contracted are also STF and thus would give $0$ on contraction with
  any non-STF part of the given quantities.}
\begin{subequations}
  \label{eq:LowestOrderMultipoles}
  \begin{align}
    \label{eq:LowestOrderMultipoleU}
    U_{jk} &\propto \nHat_{\langle j} \nHat_{k\rangle} \\
    \label{eq:LowestOrderMultipoleV}
    V_{jk} &\propto \ellHat_{\langle i} \nHat_{j\rangle}.
  \end{align}
\end{subequations}
They arise, respectively, from the familiar mass- and
current-quadrupole source moments
\begin{subequations}
  \label{eq:LowestOrderMultipoleSources}
  \begin{align}
    \label{eq:LowestOrderMassMultipoleSource}
    I_{jk} &\approx \int \rho\, x_{\langle j}\, x_{k\rangle}\, \d V, \\
    \label{eq:LowestOrderCurrentMultipoleSource}
    J_{jk} &\approx \int \rho\, x_{a} v_{b} \epsilon_{ab\langle j}\,
    x_{k\rangle}\, \d V,
  \end{align}
\end{subequations}
where $\rho$ is some effective density.  At this level of
approximation, and ignoring spin, we can think of $\rho$ as just being
the sum of a Dirac $\delta$ function for each of the two black
holes---which is how the factors of $x_{j}$ result in factors
proportional to $\nHat_{j}$, and $x_{a} v_{b} \epsilon_{abj}$ results
in a factor proportional to $\ellHat_j$.  These expressions, of
course, only give a small flavor of the very complicated expressions
necessary to calculate the waveform.

Multiplying the complexity of the waveform multipole tensors
themselves, the general contractions of either expression in
Eq.~\eqref{eq:LowestOrderMultipoles} with $\mathcal{Y}^{2,m}_{jk}$
will be quite complicated, involving numerous inner products like
$\nHat \cdot \zHat$, and so on.  For a precessing system measured in
an inertial frame, this can quickly lead to enormously complicated
expressions, even for fairly low-order harmonics~\cite{arunetal:2009}.
The obvious solution, then, is to express the harmonics in a rotating
frame $(\XHat, \YHat, \ZHat)$ that coincides with $(\nHat, \lambdaHat,
\ellHat)$ at any instant.  We refer to this as the ``co-orbital''
frame.

Using the co-orbital frame, we can return to the example of
Eq.~\eqref{eq:LowestOrderMultipoles} and see the simplification at
work.  The two factors of $U_{jk}$ now lie precisely in the
$\XHat$-$\YHat$ plane, and so this term provides nonzero contributions
to $h^{2,\pm 2}$ and $h^{2,0}$ only.  Similarly, $V_{jk}$ includes one
factor in the $\XHat$-$\YHat$ plane and one along $\ZHat$, and so this
term provides nonzero contributions to $h^{2,\pm 1}$ only.  This
separation of different components corresponding to different modes
would not occur in a frame not aligned to the orbital elements; all
modes would mix.  Indeed, we will see in the following section that
violations of exactly this separation of terms are the source of
asymmetries, when factors of the spin vectors pointing in arbitrary
directions replace the more orderly factors of $\nHat$, $\lambdaHat$,
and $\ellHat$.

Expressions for the spin terms in the gravitational-wave modes---which
are the only terms containing asymmetries---are collected in
Appendix~\ref{sec:Waveformmodeswithasymmetriccontributions}, including
both symmetric and non-symmetric contributions.  In
Appendix~\ref{sec:WaveformsForDataAnalysis}, we also exhibit formulas
for efficiently evaluating the gravitational-wave polarizations
measured by an inertial observer (such as a gravitational-wave
detector), given the waveform in any rotating frame (such as the
co-orbital frame).  This allows us to skip the step of transforming
the waveform into the inertial frame, which typically eliminates the
need to calculate the many elements\footnote{For example, there are
  \num{959} elements with $\ell \leq 8$, scaling roughly as
  $\ell^{3}$.} of the Wigner $\D^{(\ell)}$ matrices.

\subsection{Origin of asymmetry}
\label{sec:Originofasymmetry}
We now have the tools necessary to understand exactly where the
asymmetries come from.  By taking the contractions between the
tensors, Eq.~\eqref{eq:WaveformModesAsTensors} is also sometimes
written~\cite{thorne:1980, blanchet:2014} as
\begin{equation}
  \label{eq:WaveformModes}
  h^{\ell,m} \propto \left( U^{\ell,m} - \i\, V^{\ell,m} \right).
\end{equation}
$U^{\ell,m}$ and $V^{\ell,m}$ are the radiative mass- and
current-multipole modes.  These modes individually have well defined
parity behavior.  Thorne~\cite{thorne:1980} notes that the reality
condition on $h_{+}$ and $h_{-}$ implies the relations
\begin{subequations}
  \label{eq:MassAndCurrentMultipoleParity}
  \begin{align}
    U^{\ell,m} &= (-1)^{m} \bar{U}^{\ell,-m}, \\
    V^{\ell,m} &= (-1)^{m} \bar{V}^{\ell,-m}.
  \end{align}
\end{subequations}
This is an entirely immutable consequence of our choice of
spin-weighted spherical harmonics and of the fact that distances are
measured with real numbers.  On the other hand, Blanchet
\etal~\cite{blanchetetal:2008} report that
\begin{equation}
  \label{eq:hParity}
  h^{\ell,m} = (-1)^{\ell} \bar{h}^{\ell,-m}
\end{equation}
for the modes in the nonprecessing systems they treat.  Note that the
exponent here is $\ell$, rather than $m$ as in the preceding
equations, and that the latter equation implies a conjugation of the
factor of $\i$ that is explicitly present in
Eq.~\eqref{eq:WaveformModes}.  We saw in
Sec.~\ref{sec:Invariantcombinationsofmodes} that this equation is
equivalent to invariance of the system under reflection across the
$x$-$y$ plane.  It must be emphasized that, unlike
Eq.~\eqref{eq:MassAndCurrentMultipoleParity}, this relation is not an
essential truth, but merely a statement about modes in certain
systems, assuming a certain attitude of the decomposition basis.  It
implies that the amplitudes of modes with equal $\ell$ and opposite
$m$ will necessarily be equal; the modes will be symmetric because
$\abs{h^{\ell,m}} = \abs{(-1)^{\ell} \bar{h}^{\ell,-m}} =
\abs{h^{\ell,-m}}$.  Thus, understanding mode asymmetry will require
understanding why Eq.~\eqref{eq:hParity} is true for the modes of
Ref.~\cite{blanchetetal:2008}, and why it fails otherwise---or
equivalently, finding terms that are not invariant under reflection
across the $x$-$y$ plane.

By simply inserting Eqs.~\eqref{eq:WaveformModes}
and~\eqref{eq:MassAndCurrentMultipoleParity} into
Eq.~\eqref{eq:hParity}, we can easily see that they are consistent if
and only if the multipole modes satisfy
\begin{subequations}
  \begin{align}
    U^{\ell,m} &= 0 \quad \text{for odd $\ell+m$,} \\
    V^{\ell,m} &= 0 \quad \text{for even $\ell+m$.}
  \end{align}
\end{subequations}
And this is indeed the case for all such terms in nonprecessing
systems~\cite{fayeetal:2012}.  Recall, as mentioned in
Sec.~\ref{sec:Waveformsinthecoorbitalframe}, that $U^{\ell,m}$ and
$V^{\ell,m}$ can only be nonzero if the corresponding $U_{L}$ or
$V_{L}$ has $\abs{m}+2n$ tensor components in the $\nHat$-$\lambdaHat$
plane, for some natural number $n$.  This shows us how to find terms
that cause mode asymmetry: for even $\ell$, look for terms in $U_{L}$
with an \emph{odd} number of factors in the $\nHat$-$\lambdaHat$
plane, and terms in $V_{L}$ with an \emph{even} number of such
factors---and contrariwise for odd $\ell$.

The most important example comes from the lowest-order spin term,
which appears in $V_{jk}$.  As mentioned previously, this term arises
from the current-quadrupole source moment given in
Eq.~\eqref{eq:LowestOrderCurrentMultipoleSource}.  This involves the
integrand $\rho\, x_{a}\, v_{b}\, \epsilon_{abj}$, which just becomes
the \emph{orbital} angular momentum when spin is ignored.  When spin
is included, however, this factor gives rise to a term in the
\emph{spin} angular momentum.  Incorporating this effect from both
black holes, we can see that $V_{jk}$ now includes a term proportional
to $\Sigma_{\langle j} n_{k\rangle}$, where the spin vector
$\Sigma_{j}$ was given below
Eq.~\eqref{eq:RelativeAmplitudeDifference}.  When the spins are
aligned with $\ellHat$, this term is proportional to the basic
$V_{jk}$ expression in Eq.~\eqref{eq:LowestOrderMultipoleV}, having
just one factor in the $\nHat$-$\lambdaHat$ plane, and thus providing
symmetric contributions to $h^{2,\pm 1}$ only.  However, when
$\threevec{\Sigma}$ has any component in the $\nHat$-$\lambdaHat$
plane, it will behave more like $U_{jk}$ as given in
Eq.~\eqref{eq:LowestOrderMultipoleU}.  Thus, it will provide nonzero
contributions to $h^{2,\pm 2}$ and $h^{2,0}$; because of the factor of
$\i$ in Eq.~\eqref{eq:WaveformModes}, these contributions will
necessarily be asymmetric.

We can also think of this in terms of the effect on $\Sigma_{j}$ of
the parity-conjugation operator $\ParityInversion_{Z}$, which reflects
the system across the $X$-$Y$ plane, and the more familiar
parity-conjugation operator $\ParityInversion_{-}$, which reverses the
signs of all (polar) vectors.  Through unfortunate accidents of
history and dimensionality~\cite{doranlasenby:2003}, spin quantities
like $\Sigma_{j}$ are usually regarded as ``axial'' vectors, which do
not change under $\ParityInversion_{-}$, as is well known.  For spins
aligned with $\ZHat$, this invariance extends to
$\ParityInversion_{Z}$, which can also be regarded as
$\ParityInversion_{-}$ composed with a rotation through $\pi$ about
$\ZHat$.  However, if the spin has any component orthogonal to
$\ZHat$, the additional rotation imposed by $\ParityInversion_{Z}$
\emph{will} affect the direction of the spin.  Thus, when $\Sigma_{j}$
has a component in the $X$-$Y$ plane, terms like $\Sigma_{\langle j}
n_{k \rangle}$ will not be invariant under $\ParityInversion_{Z}$,
which means that Eq.~\eqref{eq:hParity} will not be true.

Interestingly, these terms also violate the standard result (valid for
nonprecessing systems in an appropriate frame) that $h^{\ell,m}
\propto \exp\{-\i\, m\, \Phi\}$, where $\Phi$ is the orbital phase.
Such phase factors usually come from contractions of the $\pm m$
factors of $(\xHat \mp \i\, \yHat)$ with the same number of factors of
$\nHat$---which is considered in the standard analysis to be rotating
in the $\xHat$-$\yHat$ plane with phase $\Phi$.  However, for these
spin terms, one factor of $\nHat$ is replaced by $\threevec{\Sigma}$,
which is fairly constant on the orbital timescale, so its contraction
with $(\XHat \mp \i\, \YHat)$ will be roughly constant in the
co-nutating frame.  Thus, the complex phase of this spin contribution
will vary with the orbital phase as $\exp\{-\i(m \mp 1)\Phi\}$.
Alternatively, in the co-orbital frame we have $\nHat=\XHat$, which
leads to trivial contractions, but $\threevec{\Sigma}$ rotates on an
orbital timescale with phase $-\Phi$, leading to an overall phase of
$\exp\{\mp \i\, \Phi\}$.  In either type of frame, the resulting phase
factor is the source of the oscillations.  And so, not only will
asymmetries generally be present in waveforms from precessing systems,
but they will oscillate on an orbital timescale, as seen in
Sec.~\ref{sec:AsymmetriesInNumericalData}.  We will now show that the
same features do indeed appear in \pN data when asymmetries are
included.

\subsection{Asymmetries in \pN data}
\label{sec:Asymmetriesinpndata}

Using the asymmetric terms described above, we can construct a \pN
waveform corresponding to the numerical data discussed in
Sec.~\ref{sec:AsymmetriesInNumericalData}, and reproduce each of the
plots given above to see if those features are also present in the \pN
waveform.  A full comparison between NR and \pN is beyond the scope of
this paper; our purpose here is simply to show that the asymmetric \pN
terms are capable of reproducing the features seen in the numerical
data.

The \pN initial parameters are chosen naively, using the horizon
quantities from the NR data measured roughly $800\,M$ after the
beginning of the simulation.  This numerical data is taken with
respect to the arbitrary coordinates in the strongly dynamical part of
the spacetime.  In particular, we do not expect the coordinates in the
vicinities of the horizons to have any clear relation to coordinates
at $\scriplus$---where the waveforms are ostensibly measured.
Therefore, we align the PN waveform to the NR waveform by shifting in
time and by rotating, to optimize the alignment between the waveform
frames~\cite{boyle:2013, boyleetal:2014b}.

%%%%%%%%%%%%%%%%%%%%%%%%%%%%%%%%%%%%%%%%%%%%%%%%%%%%%%%%%%%%%%%%%%%%%%
\begin{figure*}
  % \tikzset{external/remake next=true}
  \includegraphics{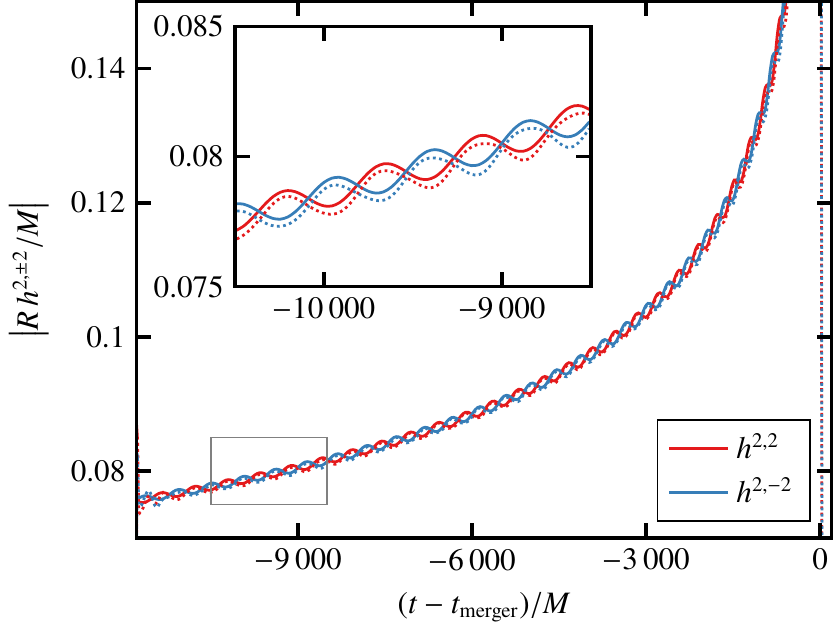} %
  \hfill
  % \tikzset{external/remake next=true}
  \includegraphics{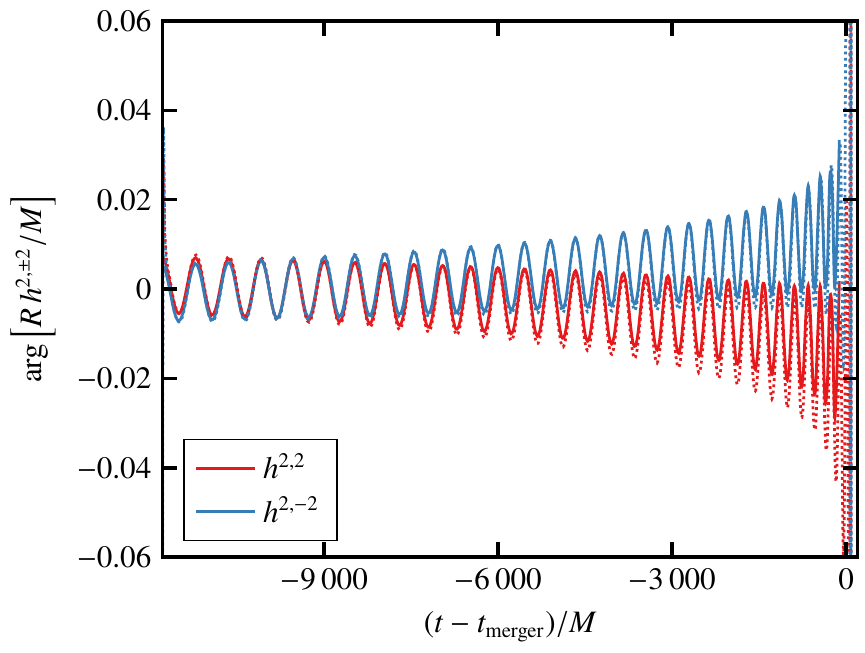} %
  \\
  % \tikzset{external/remake next=true}
  \includegraphics{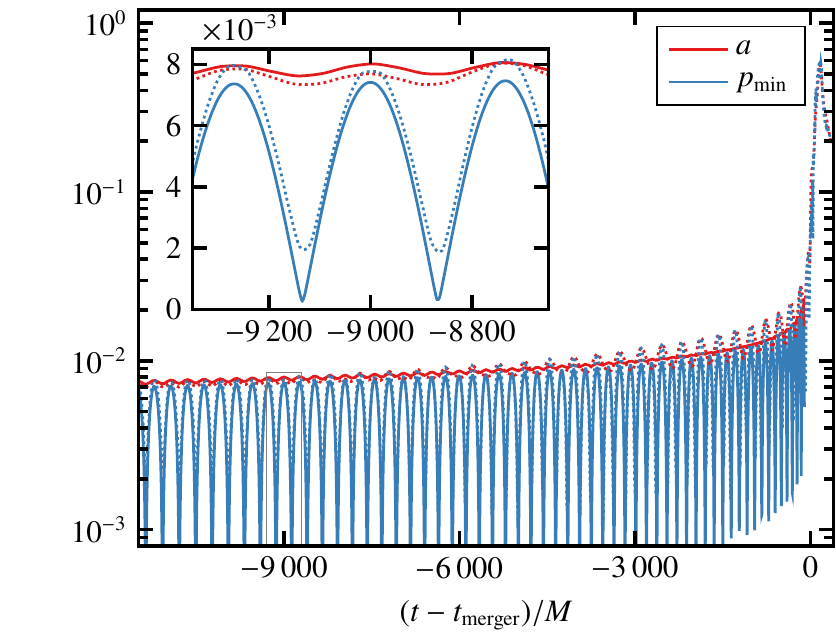} %
  \hfill
  % \tikzset{external/remake next=true}
  \includegraphics{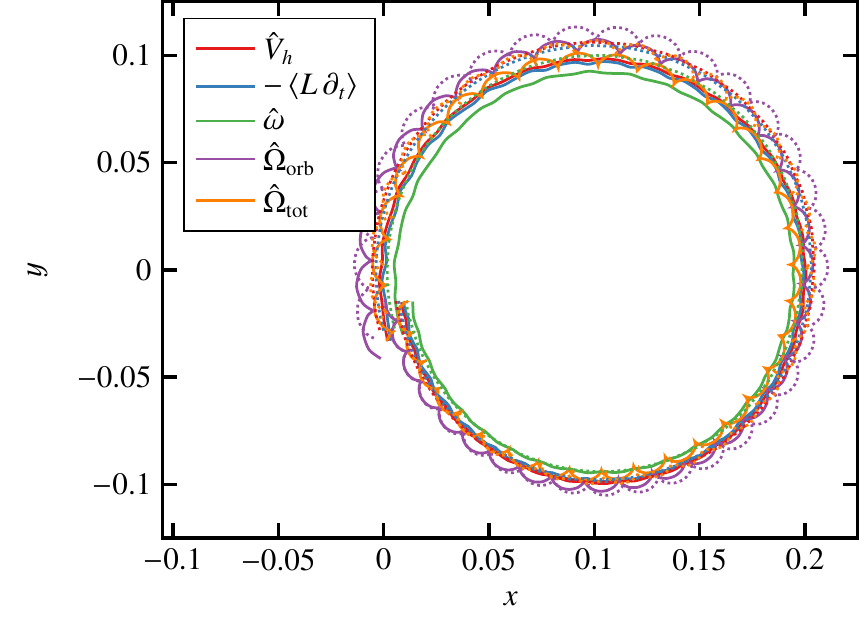} %
  \\
  \caption{ \label{fig:PNComparisons} %
    \CapName{Comparing asymmetries in the NR and \pN data} These plots
    reproduce the preceding figures, but now show the \pN data as
    solid curves, and the corresponding NR data as dotted curves.  We
    see imperfect quantitative agreement, as must be expected of
    approximate analytical waveforms.  For example, the overall \pN
    amplitude is too large compared to NR (top left).  Similarly, the
    \pN $\NormalizedParityViolation$ is smaller than expected, while
    the $\NormalizedAntisymmetry$ is larger than expected (bottom
    left).  Nonetheless, the qualitative agreement is impressive,
    showing that each of the features described in the numerical data
    [Sec.~\ref{sec:AsymmetriesInNumericalData}] is also present in the
    \pN waveforms once the antisymmetric terms have been included.
  }
\end{figure*}
%%%%%%%%%%%%%%%%%%%%%%%%%%%%%%%%%%%%%%%%%%%%%%%%%%%%%%%%%%%%%%%%%%%%%%
In Fig.~\ref{fig:PNComparisons}, we reproduce each of the plots from
Sec.~\ref{sec:AsymmetriesInNumericalData}, but use \pN data.  For
comparison, the NR data are included as dotted curves of corresponding
colors.  Broadly speaking, the \pN and NR results are qualitatively
very similar.  The antisymmetry and parity violation would be zero
without the terms described above.  By including them, we obtain
features much like those seen in the numerical data.  There are, of
course, quantitative disagreements.  For example, the overall \pN
amplitudes of the $h^{2,\pm2}$ modes are larger than the NR
amplitudes, while the size of the oscillations is very nearly correct
(in both amplitude and phase).  Similarly, while the vectors derived
from the waveform precess far more smoothly than the nutating orbital
vectors, there are larger oscillations present in the \pN data than in
the NR data.  Of course, such disagreements are to be expected from
the approximate formulas of \pN.  The important point is that the
agreement is drastically improved when antisymmetric terms are
included in the \pN formulas.

%%%%%%%%%%%%%%%%%%%%%%%%%%%%%%%%%%%%%%%%%%%%%%%%%%%%%%%%%%%%%%%%%%%%%%
%%%%%%%%%%%%%%%%%%%%%%%%%%%%%%%%%%%%%%%%%%%%%%%%%%%%%%%%%%%%%%%%%%%%%%
\section{Antisymmetries in binary recoil}
\label{sec:Recoil}
One of the most consequential discoveries of the era of numerical
relativity was the discovery of
``super-kicks''~\cite{campanellietal:2007a, campanellietal:2007b,
  tichymarronetti:2007, gonzalezetal:2007, brugmannetal:2008,
  loustozlochower:2011}, in which the linear momentum carried off by
gravitational waves from binaries with spins in the orbital plane can
be so large that the merged system is left with a very large recoil
velocity---possibly in excess of \SI{5000}{\kilo\meter \per
  \second}~\cite{loustozlochower:2011}.  Linear momentum can only be
carried off if the distribution of energy in the radiated waves is
asymmetric.  Our study of the asymmetry in gravitational radiation can
therefore improve our understanding of the origin of the recoil.

Translating Thorne's Eq.~(4.18)~\cite{thorne:1980} into our notation,
we have the following expression for the linear-momentum flux in the
form of gravitational waves:
\begin{equation}
  \label{eq:LinearMomentumFlux}
  \frac{\d\threevec{p}}{\d\Omega\, \d t} =
  \frac{\LuminosityDistance^{2}} {16\,\pi}\, \abs{\frac{\d h}{\d
      t}}^{2} \, \rHat,
\end{equation}
where $\rHat$ is the direction from the source to the point in
question and $\LuminosityDistance$ is the distance from the source to
the observation sphere (so that $\LuminosityDistance\, h$ asymptotes
to a nonzero constant).  To find the total linear-momentum emission,
we can integrate over all angles, expand the integrand in terms of the
$h^{\ell,m}$ modes, and use the fact that each component of $\rHat$
may be written as a sum of $\ell=1$ spherical harmonics.  Defining the
modes $\rHat_{j}^{\ell,m}$ as implied by
\begin{equation}
  \label{eq:nHatModes}
  \rHat_{j} = \sqrt{\frac{2\pi}{3}}\, \left( Y_{1,-1} - Y_{1,1},
    iY_{1,-1} + iY_{1,1}, \sqrt{2}Y_{1,0} \right)_{j},
\end{equation}
we can do the integral explicitly, using a formula from
Ref.~\cite{*[{Eq.~(C.7) of }] [{.}] alcubierre:2008}, and find
\begin{widetext}
  \begin{subequations}
    \label{eq:MomentumEmission}
    \begin{equation}
      \label{eq:MomentumEmission_general}
      \frac{\d p_j}{\d t} = \frac{\LuminosityDistance^{2}}{16\pi}\,
      \sum_{\ell,\ell',m,m'} \rHat_j^{1,m'-m} \dot{h}^{\ell,m}\,
      \bar{\dot{h}}^{\ell',m'} (-1)^{m'}\,
      \sqrt{\frac{3(2\ell+1)(2\ell'+1)}{4\pi}}
      \begin{pmatrix}
        \ell & \ell' & 1\\
        m & -m' & m'-m
      \end{pmatrix}
      \begin{pmatrix}
        \ell & \ell' & 1\\
        2 & -2 & 0
      \end{pmatrix}.
    \end{equation}
    We use $\dot{p}_{j} \defined \d p_{j}/\d t$ and $\dot{h}
    \defined \d h/\d t$ for brevity.  For the particularly interesting
    case of the $z$ component, this reduces to
    \begin{equation}
      \label{eq:MomentumEmission_z}
      \dot{p}_z = \frac{\LuminosityDistance^{2}}{16\pi}\,
      \sum_{\ell,\ell',m} \dot{h}^{\ell,m}\, \bar{\dot{h}}^{\ell',m}
      (-1)^{m}\, \sqrt{(2\ell+1)(2\ell'+1)}
      \begin{pmatrix}
        \ell & \ell' & 1\\
        m & -m & 0
      \end{pmatrix}
      \begin{pmatrix}
        \ell & \ell' & 1\\
        2 & -2 & 0
      \end{pmatrix}.
    \end{equation}
  \end{subequations}
\end{widetext}
In each of these equations, the last two factors are Wigner's 3$j$
symbols.  Noting various properties of those symbols, we can see that
the sums over $\ell'$ only run over $\{\ell-1,\ell,\ell+1\}$.  We also
know that the sum over $m'$ in Eq.~\eqref{eq:MomentumEmission_general}
runs over $\{m-1,m+1\}$ for the $x$ and $y$ components, and reduces to
$m'=0$ for the $z$ component.  These facts limit mode mixing, and thus
make the sum far more tractable.  These expressions are independent of
any \pN expansion, and are given directly in terms of the waveform
modes, rather than the source multipoles.

For example, we can find all terms in
Eq.~\eqref{eq:MomentumEmission_z} to which the $\dot{h}^{2,\pm2}$
modes contribute:
\begin{multline}
  \dot{p}_z = \frac{\LuminosityDistance^{2}} {12\,\pi}\,
  \left(\abs{\dot{h}^{2,2}}^2 - \abs{\dot{h}^{2,-2}}^{2} \right) \\ +
  \frac{\LuminosityDistance^{2}} {24\,\pi}\, \sqrt{\frac{5}{7}}\, \Re
  \left[ \dot{h}^{2,2}\, \bar{\dot{h}}^{3,2} + \dot{h}^{2,-2}\,
    \bar{\dot{h}}^{3,-2} \right] + \ldots,
\end{multline}
where remaining terms do not involve $\dot{h}^{2,\pm2}$.  This
equation is entirely general, independent of the frame used to
decompose the waveform into modes and of any \pN or other
approximations.  However, this result is particularly interesting in
the co-orbital frame, because it expresses the linear momentum emitted
by gravitational waves in a direction orthogonal to the orbital plane;
this is the origin of ``super-kicks''.  The first term involves
exactly the asymmetry visible in Figs.~\ref{fig:ModeAsymmetries}
and~\ref{fig:Asymmetry}.  Similarly, the second term would be zero if
the $h^{2,\pm2}$ and $h^{3,\pm2}$ modes obeyed the parity-invariance
equation, Eq.~\eqref{eq:ModeSymmetry}.\footnote{This is true for the
  $z$ component of the momentum only.  Similar terms appear in the
  expressions for the $x$ and $y$ components, and can be nonzero
  without violating Eq.~\eqref{eq:ModeSymmetry}.  Those terms are the
  mechanism for emission of linear momentum from nonprecessing
  systems.}

The obvious parity violation evident in
Eq.~\eqref{eq:MomentumEmission_z} might tempt us to conjecture that
only the antisymmetric portions of the waveform are involved in
producing a recoil.  In fact, we can show that this is not at all the
case.  To begin, we consider $\dot{p}_{j}$ to be a functional
operating on the waveform: $\dot{p}_{j}[h]$.  We also define $P_{k}$
to be any of the three involution operators $(P_{x}, P_{y}, P_{z})$.
Using this notation, we expect on physical grounds to find
\begin{equation}
  \label{eq:InvolutionsOfMomentumFlux}
  \ParityInversion_{k} \dot{p}_{j} \left[ h \right] = \dot{p}_{j}
  \left[ \ParityInversion_{k} h \right].
\end{equation}
That is, any parity inversion of the recoil produced by $h$ is the
same as the recoil produced by that parity inversion of $h$.  Though
the physical interpretation is clear, it is not immediately obvious
that our mathematical expressions for the recoil behave correctly
under parity inversions.  However, if we consider the integral of
Eq.~\eqref{eq:LinearMomentumFlux} needed to find $\ParityInversion_{k}
\dot{p}_{j} \left[ h \right]$, we can use $P_{k} \rHat = -\rHat$ and
perform a change of variables, for which the Jacobian determinant is
$-1$, leaving us with an expression identical to that of $\dot{p}_{j}
\left[ \ParityInversion_{k} h \right]$~\cite{frankel:2004}.  We can
also verify this fact for the particular expressions in
Eqs.~\eqref{eq:MomentumEmission} by inserting the various
transformation formulas given in
Eqs.~\eqref{eq:ModeTransformationLaws}, relabeling the summation
indices, and using properties of the Wigner $3j$ symbols.

As in Sec.~\ref{sec:Invariantcombinationsofmodes}, we define the
projection operators $\ParityInvariantProjection_{k} \defined
\frac{1}{2} (1 - \ParityInversion_{k})$, which retain only the
portions of the waveform that reverse sign under
$\ParityInversion_{k}$.  We also introduce the complementary
projection operators $\ParityVariantProjection_{k} \defined
\frac{1}{2} (1 + \ParityInversion_{k})$, which retain only the
portions of the waveform that do not change under
$\ParityInversion_{k}$.  It is not hard to see that these projection
operators result in eigenfunctions of the corresponding parity
inversions:
\begin{equation}
  \label{eq:ParityInversionOfProjections}
  \ParityInversion_{k} \ParityInvariantProjection_{k} =
  -\ParityInvariantProjection_{k} \qquad \text{and} \qquad
  \ParityInversion_{k} \ParityVariantProjection_{k}
  = \ParityVariantProjection_{k}.
\end{equation}
Now, using Eq.~\eqref{eq:InvolutionsOfMomentumFlux}, we have
\begin{equation}
  \ParityInversion_{k} \dot{p}_{j} [\ParityVariantProjection_{k} h] =
  \dot{p}_{j} [\ParityInversion_{k}\ParityVariantProjection_{k} h] =
  \dot{p}_{j} [\ParityVariantProjection_{k} h].
\end{equation}
This says that the parity inverse of the recoil vector equals itself.
When $j \neq k$, this doesn't tell us anything.  However, when $j=k$,
the sign of that component of the vector must reverse under parity
inversion.  Thus, we have
\begin{equation}
  \label{eq:ParitySymmetricMomentumEmission}
  \dot{p}_{j} [\ParityVariantProjection_{j} h] = 0.
\end{equation}
The same logic, using $\dot{p}_{j}[-h] = \dot{p}_{j}[h]$, shows that
\begin{equation}
  \label{eq:ParityAntisymmetricMomentumEmission}
  \dot{p}_{j} [\ParityInvariantProjection_{j} h] = 0.
\end{equation}
Taken together, we can interpret these equations to say that recoil in
a given direction requires parity-violating asymmetry in that
direction, but recoil is not caused by the antisymmetric part alone;
it is caused by the \emph{interaction} of the parity-violating and
parity-satisfying parts of the waveform.  In fact, we can even rewrite
Eq.~\eqref{eq:LinearMomentumFlux} as
\begin{equation}
  \label{eq:LinearMomentumFluxByProjections}
  \frac{\d\threevec{p}}{\d t} = \frac{\LuminosityDistance^{2}}
  {8\,\pi}\, \int \Re \left\{ \ParityInvariantProjection_{j} \dot{h}\,
    \overline{\ParityVariantProjection_{j} \dot{h}} \right\} \,
  \rHat\, \d\Omega,
\end{equation}
which is valid for any choice of $j$, and shows explicitly that the
net recoil is a product of the symmetric and antisymmetric parts of
the waveform.

%%%%%%%%%%%%%%%%%%%%%%%%%%%%%%%%%%%%%%%%%%%%%%%%%%%%%%%%%%%%%%%%%%%%%%
%%%%%%%%%%%%%%%%%%%%%%%%%%%%%%%%%%%%%%%%%%%%%%%%%%%%%%%%%%%%%%%%%%%%%%
\section{Conclusions}
\label{sec:Conclusions}

We have shown that gravitational waves from precessing black-hole
binaries include features that are inherently asymmetric, in the sense
that no rotation can eliminate them.  The asymmetries are caused by
the presence of spin components that are not aligned with the symmetry
axis of the orbital motion.  The effects on the waveforms can be very
large, causing direction-dependent effects on the relative amplitude
of greater than \SI{50}{\percent} at merger.  These asymmetries can
already be found in some \pN expressions for the metric
perturbation---though, in the literature, the expressions are
generally simplified to eliminate asymmetries before being translated
into modes.  We separated the symmetric and anti-symmetric components
of the waveform, and showed that neither component alone is
responsible for binary recoil.  Rather, binary recoil is a result of
the interaction between the completely symmetric and the completely
anti-symmetric components of the gravitational radiation.

We showed that the co-orbital frame significantly simplifies
expressions for the \pN waveform modes, without any approximations of
small precession angles, \etc.  Because the attitude of the co-orbital
frame must be calculated for any precessing system, there is no
computational overhead in this approach---only savings from the
simplified expressions.  By computing the symmetric and anti-symmetric
components of the waveforms separately, we can regain the efficiency
of simultaneously calculating the $h^{\ell,m}$ and $h^{\ell,-m}$ mode
pairs, which is used for nonprecessing systems [see
Appendix~\ref{sec:Waveformmodeswithasymmetriccontributions}].
Finally, as demonstrated in
Appendix~\ref{sec:WaveformsForDataAnalysis}, it is possible to use
modes expressed in the co-orbital frame (or any rotating frame)
\emph{directly} in calculation of the waveform observed by an inertial
detector, rather than going through the computationally burdensome
step of rotating the modes.

We suggest, therefore, that analyzing analytical mode expressions in
the co-orbital frame and separating the symmetric from the
anti-symmetric components lead to improvements in both the analytical
treatment of waveforms and the numerical performance of related
calculations.  This will be important for gravitational-wave
astronomy, so that these fascinating sources may be studied in greater
detail and with higher accuracy.  Precession imprints information on
the gravitational-wave signal, which can potentially allow measurement
of otherwise ambiguous features of astrophysical sources, and thus
improve the scientific output of gravitational-wave
astronomy~\cite{LVC:2013}.

Future work will be needed to improve the coverage of \pN terms for
precessing waveforms, and to draw numerical and analytical work
closer.  In particular, more extensive comparisons of \pN and NR
predictions are needed.  The rough results shown in
Sec.~\ref{sec:Asymmetriesinpndata} are only meant to be qualitative;
they are tainted most prominently by the naive method used to
determine the parameters of the \pN system from coordinate-dependent
quantities in the numerical data.

All of the methods and results of this paper are included as computer
code in the \href{\ancillaryURL} {ancillary files} on this paper's
arXiv page.  These include the expressions for the waveform modes,
code to evolve precessing binaries and compute the modes in the
co-orbital frame, and code to transform and evaluate those modes.  We
hope that this may form a basis for future work.

% \begin{itemize}
%  \item Evaluate at a point efficiently.
%  \item Transform from co-orbital to co-rotating directly
%  \item Eventually, look into ``resummed'' or ``refactorized''
%   versions of \pN, in which $M/r$, $\dot{r}$, \etc, are kept around,
%   rather than re-expanded and truncated.  Also, think about the flux
%   expression.
% \end{itemize}

%%%%%%%%%%%%%%%%%%%%%%%%%%%%%%%%%%%%%%%%%%%%%%%%%%%%%%%%%%%%%%%%%%%%%%
\begin{acknowledgments}
  It is our pleasure to thank Jeandrew Brink, Geoffrey Lovelace, and
  Saul Teukolsky for useful conversations.  This project was supported
  in part by the Sherman Fairchild Foundation; by NSF Grants No.\
  PHY-1306125 and AST-1333129; and
  by NSERC of Canada and the Canada Research Chairs Program.

  The computations presented in this paper were performed on the
  \texttt{Zwicky} cluster hosted at Caltech by the Center for Advanced
  Computing Research, which was funded by the Sherman Fairchild
  Foundation and the NSF MRI-R\textsuperscript{2} program; and on the
  \texttt{GPC} and \texttt{Gravity} clusters at the SciNet HPC
  Consortium~\cite{scinet}, funded by: the Canada Foundation for
  Innovation under the auspices of Compute Canada; the Government of
  Ontario; Ontario Research Fund--Research Excellence; and the
  University of Toronto.
\end{acknowledgments}

%%%%%%%%%%%%%%%%%%%%%%%%%%%%%%%%%%%%%%%%%%%%%%%%%%%%%%%%%%%%%%%%%%%%%%
%%%%%%%%%%%%%%%%%%%%%%%%%%%%%%%%%%%%%%%%%%%%%%%%%%%%%%%%%%%%%%%%%%%%%%
% \appendix* % Use \appendix* if there is just one appendix
\appendix % Use \appendix if there are multiple appendices

\section{Waveform modes with asymmetric contributions}
\label{sec:Waveformmodeswithasymmetriccontributions}

We now give the formulas for the contributions to the waveform modes
from spins in the co-orbital frame.  This form of the modes is useful
as it is particularly simple; it factors out the dependence on
attitude of the precessing binary including all orbital motion.  Also,
we can simply add these expressions directly to the equivalent
non-spin mode contributions, to achieve more complete and accurate
results.  The combination may not be at a consistent \pN order---since
terms with spin will only be kept up to 2\pN, while terms without may
be kept up to 3.5\pN---but the results will be numerically accurate,
and will smoothly transition from nonprecessing to strongly
precessing.  Finally, calculating these modes numerically is
efficient, as the symmetric and antisymmetric parts may be calculated
separately, and then added with appropriate signs and conjugations to
give both $h^{\ell,m}$ and $h^{\ell,-m}$, rather than calculating each
individually.

We begin by defining the various elements.  The black holes have
masses $M_{1}$ and $M_{2}$, and spins $\vec{S}_{1}$ and $\vec{S}_{2}$,
respectively.  The additional symbols we will use in writing the modes
are
\begin{subequations}
  \label{eq:SymbolDefinitions}
  \begin{gather}
    M \defined M_1 + M_2, \\
    \nu \defined \frac{M_{1}\, M_{2}} {M^{2}}, \\
    \delta \defined \frac{M_{1}-M_{2}} {M}, \\
    \vec{S} \defined \vec{S}_{1} + \vec{S}_{2}, \\
    \vec{\Sigma} \defined M \left( \frac{\vec{S}_{2}}{M_{2}} -
      \frac{\vec{S}_{1}}{M_{1}} \right), \\
    v \defined \left( M\, \abs{\OmegaOrb} \right)^{1/3} = \sqrt{x}.
  \end{gather}
\end{subequations}
The last symbol is the usual \pN-expansion parameter, and we have set
$G = c = 1$.  Again, we note that $\nHat$ is a unit vector pointing
\emph{from} black hole $2$ \emph{to} black hole $1$; $\lambdaHat$ is a
unit vector in the direction of $\frac{\d}{\d t} \nHat$; and $\ellHat
= \nHat \times \lambdaHat$.

The modes are derived from expressions for the metric perturbation
$h_{jk}$, as given by Eqs.~(4.9) of Ref.~\cite{kidder:1995} and
Eqs.~(4.13) and~(4.15) of Ref.~\cite{buonannoetal:2012}.  These
include the spin-orbit, $\text{spin}_{1}$-$\text{spin}_{2}$,
$\text{spin}_{1}^{2}$, and $\text{spin}_{2}^{2}$ terms through 2\pN
order, relative to the leading-order (non-spin) term in the metric
perturbation.  Using the substitutions $(\nHat, \lambdaHat, \ellHat)
\mapsto (\xHat, \yHat, \zHat)$, the complex $h$ field is derived from
$h_{jk}$ as
\begin{equation}
  \label{eq:Complexh}
  h \defined \frac{1}{2} \left[ \left( \varphi^{j} \varphi^{k} -
      \vartheta^{j} \vartheta^{k} \right) + \i\, \left( \vartheta^{j}
      \varphi^{k} + \varphi^{j} \vartheta^{k} \right) \right] h_{jk},
\end{equation}
where $\vartheta$ and $\varphi$ are the usual spherical coordinates of
the $(\nHat, \lambdaHat, \ellHat) = (\xHat, \yHat, \zHat)$ system.
Equation~\eqref{eq:Complexh} expresses $h = h_{+} - \i h_{\times}$, so
$h$ is a field of spin weight $s=-2$.  Note that there are many subtly
different conventions throughout the literature for the various
quantities we treat here.  Our choices are internally consistent, and
made so that the results for the modes given in Eq.~\eqref{eq:h_spin}
are consistent with other results \emph{for gravitational-wave modes}
in the literature---specifically the review in
Ref.~\cite{blanchet:2014}.  Note in particular the relation between
the field point $\varphi$ in our co-orbital frame and the orbital
phase $\Phi$ in the more common frame for which the system orbits in
the $x$-$y$ plane while the field point is in the $y$-$z$ plane:
formulas given in the two frames can be equated when we define
$\varphi = \pi/2-\Phi$.

We then find the SWSH mode decomposition according to
\begin{equation}
  \label{eq:SWSHDecomposition}
  h^{\ell,m} \defined \int_{0}^{\pi} \int_{0}^{2\pi} h(\vartheta,
  \varphi)\, \mTwoYbarlm{\ell,m}(\vartheta, \varphi)\, \d\varphi\,
  \sin\vartheta\, \d\vartheta.
\end{equation}
As is standard, we rescale all modes by the leading-order (non-spin)
term in $h^{2,2}$; that is, we present modes of the rescaled field
\begin{equation}
  \label{eq:hHat}
  \hat{h} \defined \frac{1}{8\nu v^{2}} \sqrt{\frac{5}{\pi}}\,
  \frac{\LuminosityDistance}{M}\, h.
\end{equation}
The results are the modes in the co-orbital frame [see
Sec.~\ref{sec:Waveformsinthecoorbitalframe}].  These modes can, of
course, be rotated into any other frame (including inertial and
co-rotating frames) using Eq.~\eqref{eq:ModeRotation}.

For presentation purposes, we display each term in the form
$\{\ParityVariantProjection_{z} h\} + \{\ParityInvariantProjection_{z}
h\}$; the first brace group is symmetric (does not change sign) under
reflection across the orbital plane, while the second is
antisymmetric.  Of course, these parts need not be recalculated for
each mode.  The terms may be calculated separately for modes with
positive $m$, and the modes with corresponding $-m$ value constructed
simply by combining the two types of terms after the appropriate
conjugations and sign changes.  We include the symmetric contributions
to the modes coming from spin for completeness, even though they are
essentially available from other sources~\cite{ninja2:2012}.  The full
expressions, including contributions from non-spin parts, are given in
Mathematica and IPython notebooks available in the
\href{\ancillaryURL} {ancillary materials} on this paper's arXiv page,
along with code that computes the \pN trajectories and produces the
full \pN waveform.

\begingroup
\allowdisplaybreaks
\begin{widetext}
  \begin{subequations}
    \label{eq:h_spin}
    \begin{align}
    \label{eq:h_spin_2_m2}%
    \hat{h}_{\text{spin}}^{2,-2}
    &= \left\{-\frac{2 v^3 \left(3 S_{\ell}+\delta \Sigma_{\ell}\right)}{3 M^2}+\frac{v^4 \left(-22 {S_1}_n {S_2}_n+15 i {S_1}_{\lambda} {S_2}_n+12 {S_1}_{\ell} {S_2}_{\ell}+15 i {S_1}_n {S_2}_{\lambda}+10 {S_1}_{\lambda} {S_2}_{\lambda}\right)}{6 M^4 \nu}\right\} \nonumber \\
    &\quad+ \left\{\frac{v^2 \left(\Sigma_{\lambda}-i \Sigma_n\right)}{2 M^2}+\frac{v^4 \left[182 i \delta S_n-19 \delta S_{\lambda}+14 i (7-20 \nu) \Sigma_n-(5-43 \nu) \Sigma_{\lambda}\right]}{84 M^2}\right\}
    \\
    \label{eq:h_spin_2_m1}%
    \hat{h}_{\text{spin}}^{2,-1}
    &= \left\{-\frac{i v^2 \Sigma_{\ell}}{2 M^2}+\frac{i v^4 \left[86 \delta S_{\ell}+(79-139 \nu) \Sigma_{\ell}\right]}{42 M^2}\right\} \nonumber \\
    &\quad+ \left\{\frac{v^3 \left(-25 S_n+4 i S_{\lambda}-13 \delta \Sigma_n+4 i \delta \Sigma_{\lambda}\right)}{6 M^2}+\frac{3 v^4 \left({S_1}_{\ell} {S_2}_n+{S_1}_n {S_2}_{\ell}\right)}{2 M^4 \nu}\right\}
    \\
    \label{eq:h_spin_2_0}%
    \hat{h}_{\text{spin}}^{2,0}
    &= \left\{\frac{\sqrt{\frac{2}{3}} v^4 \left({S_1}_n {S_2}_n-{S_1}_{\lambda} {S_2}_{\lambda}\right)}{M^4 \nu}\right\} + \left\{\frac{i v^2 \Sigma_n}{\sqrt{6} M^2}+\frac{i v^4 \left[255 \delta S_n+(45-506 \nu) \Sigma_n\right]}{21 \sqrt{6} M^2}\right\}
    \\
    \label{eq:h_spin_2_1}%
    \hat{h}_{\text{spin}}^{2,1}
    &= \left\{\frac{i v^2 \Sigma_{\ell}}{2 M^2}-\frac{i v^4 \left[86 \delta S_{\ell}+(79-139 \nu) \Sigma_{\ell}\right]}{42 M^2}\right\} \nonumber \\
    &\quad+ \left\{\frac{v^3 \left(25 S_n+4 i S_{\lambda}+13 \delta \Sigma_n+4 i \delta \Sigma_{\lambda}\right)}{6 M^2}-\frac{3 v^4 \left({S_1}_{\ell} {S_2}_n+{S_1}_n {S_2}_{\ell}\right)}{2 M^4 \nu}\right\}
    \\
    \label{eq:h_spin_2_2}%
    \hat{h}_{\text{spin}}^{2,2}
    &= \left\{-\frac{2 v^3 \left(3 S_{\ell}+\delta \Sigma_{\ell}\right)}{3 M^2}+\frac{v^4 \left(-22 {S_1}_n {S_2}_n-15 i {S_1}_{\lambda} {S_2}_n+12 {S_1}_{\ell} {S_2}_{\ell}-15 i {S_1}_n {S_2}_{\lambda}+10 {S_1}_{\lambda} {S_2}_{\lambda}\right)}{6 M^4 \nu}\right\} \nonumber \\
    &\quad+ \left\{-\frac{v^2 \left(\Sigma_{\lambda}+i \Sigma_n\right)}{2 M^2}+\frac{v^4 \left[182 i \delta S_n+19 \delta S_{\lambda}+14 i(7-20 \nu)\Sigma_n+(5-43 \nu)\Sigma_{\lambda}\right]}{84 M^2}\right\}
    \\
    \label{eq:h_spin_3_m3}%
    \hat{h}_{\text{spin}}^{3,-3}
    &= \left\{\frac{3 i \sqrt{\frac{15}{14}} v^4 \left[7 \delta S_{\ell}+(3-9 \nu) \Sigma_{\ell}\right]}{8 M^2}\right\} + \left\{-\frac{\sqrt{\frac{10}{21}} v^3 \left[S_n+iS_{\lambda}+\delta \left(\Sigma_n+i\Sigma_{\lambda}\right)\right]}{M^2}\right\}
    \\
    \label{eq:h_spin_3_m2}%
    \hat{h}_{\text{spin}}^{3,-2}
    &= \left\{-\frac{2 \sqrt{\frac{5}{7}} v^3 \left(S_{\ell}+\delta \Sigma_{\ell}\right)}{3 M^2}\right\} + \left\{\frac{\sqrt{\frac{5}{7}} v^4 \left[25 \delta \left(4 i S_n+S_{\lambda}\right)+4 i (13-55 \nu) \Sigma_n+(17-83 \nu) \Sigma_{\lambda}\right]}{24 M^2}\right\}
    \\
    \label{eq:h_spin_3_m1}%
    \hat{h}_{\text{spin}}^{3,-1}
    &= \left\{\frac{i v^4 \left[\delta S_{\ell}+(5-15 \nu) \Sigma_{\ell}\right]}{24 \sqrt{14} M^2}\right\} + \left\{\frac{\sqrt{\frac{2}{7}} v^3 \left[S_n-i S_{\lambda}+\delta \left(\Sigma_n-i \Sigma_{\lambda}\right)\right]}{3 M^2}\right\}
    \\
    \label{eq:h_spin_3_0}%
    \hat{h}_{\text{spin}}^{3,0}
    &= \left\{0\right\} + \left\{-\frac{v^4 \left[17 \delta S_{\lambda}+(9-35 \nu) \Sigma_{\lambda}\right]}{4 \sqrt{42} M^2}\right\}
    \\
    \label{eq:h_spin_3_1}%
    \hat{h}_{\text{spin}}^{3,1}
    &= \left\{\frac{i v^4 \left[\delta S_{\ell}+(5-15 \nu) \Sigma_{\ell}\right]}{24 \sqrt{14} M^2}\right\} + \left\{\frac{\sqrt{\frac{2}{7}} v^3 \left[S_n+i S_{\lambda}+\delta \left(\Sigma_n+i \Sigma_{\lambda}\right)\right]}{3 M^2}\right\}
    \\
    \label{eq:h_spin_3_2}%
    \hat{h}_{\text{spin}}^{3,2}
    &= \left\{\frac{2 \sqrt{\frac{5}{7}} v^3 \left(S_{\ell}+\delta \Sigma_{\ell}\right)}{3 M^2}\right\} + \left\{\frac{\sqrt{\frac{5}{7}} v^4 \left[25 \delta \left(-4 i S_n+S_{\lambda}\right)-4 i (13-55 \nu) \Sigma_n+(17-83 \nu) \Sigma_{\lambda}\right]}{24 M^2}\right\}
    \\
    \label{eq:h_spin_3_3}%
    \hat{h}_{\text{spin}}^{3,3}
    &= \left\{\frac{3 i \sqrt{\frac{15}{14}} v^4 \left[7 \delta S_{\ell}+(3-9 \nu) \Sigma_{\ell}\right]}{8 M^2}\right\} + \left\{-\frac{\sqrt{\frac{10}{21}} v^3 \left[S_n-i S_{\lambda}+\delta \left(\Sigma_n-i \Sigma_{\lambda}\right)\right]}{M^2}\right\}
    \\
    \label{eq:h_spin_4_m4}%
    \hat{h}_{\text{spin}}^{4,-4}
    &= \left\{0\right\} + \left\{-\frac{9 \sqrt{\frac{5}{7}} v^4 \left[\delta \left(S_{\lambda}-i S_n\right)+(1-3 \nu) \left(\Sigma_{\lambda}-i \Sigma_n\right)\right]}{8 M^2}\right\}
    \\
    \label{eq:h_spin_4_m3}%
    \hat{h}_{\text{spin}}^{4,-3}
    &= \left\{\frac{9 i \sqrt{\frac{5}{14}} v^4 \left[\delta S_{\ell}+(1-3 \nu) \Sigma_{\ell}\right]}{8 M^2}\right\} + \left\{0\right\}
    \\
    \label{eq:h_spin_4_m2}%
    \hat{h}_{\text{spin}}^{4,-2}
    &= \left\{0\right\} + \left\{-\frac{\sqrt{5} v^4 \left[\delta \left(13 S_{\lambda}+14 i S_n\right)+(1-3 \nu)  (13 \Sigma_{\lambda}+14 i \Sigma_n)\right]}{168 M^2}\right\}
    \\
    \label{eq:h_spin_4_m1}%
    \hat{h}_{\text{spin}}^{4,-1}
    &= \left\{-\frac{i \sqrt{\frac{5}{2}} v^4 \left[\delta S_{\ell}+(1-3 \nu) \Sigma_{\ell}\right]}{168 M^2}\right\} + \left\{0\right\}
    \\
    \label{eq:h_spin_4_0}%
    \hat{h}_{\text{spin}}^{4,0}
    &= \left\{0\right\} + \left\{\frac{i v^4 \left[\delta S_n+(1-3 \nu) \Sigma_n\right]}{84 \sqrt{2} M^2}\right\}
    \\
    \label{eq:h_spin_4_1}%
    \hat{h}_{\text{spin}}^{4,1}
    &= \left\{\frac{i \sqrt{\frac{5}{2}} v^4 \left[\delta S_{\ell}+(1-3 \nu) \Sigma_{\ell}\right]}{168 M^2}\right\} + \left\{0\right\}
    \\
    \label{eq:h_spin_4_2}%
    \hat{h}_{\text{spin}}^{4,2}
    &= \left\{0\right\} + \left\{\frac{\sqrt{5} v^4 \left[\delta \left(13 S_{\lambda}-14 i S_n\right)+(1-3 \nu) \left(13 \Sigma_{\lambda}-14 i \Sigma_n\right)\right]}{168 M^2}\right\}
    \\
    \label{eq:h_spin_4_3}%
    \hat{h}_{\text{spin}}^{4,3}
    &= \left\{-\frac{9 i \sqrt{\frac{5}{14}} v^4 \left[\delta S_{\ell}+(1-3 \nu) \Sigma_{\ell}\right]}{8 M^2}\right\} + \left\{0\right\}
    \\
    \label{eq:h_spin_4_4}%
    \hat{h}_{\text{spin}}^{4,4}
    &= \left\{0\right\} + \left\{\frac{9 \sqrt{\frac{5}{7}} v^4 \left[\delta \left(S_{\lambda}+i S_n\right)+(1-3 \nu) (\Sigma_{\lambda}+i \Sigma_n)\right]}{8 M^2}\right\}
    \end{align}
  \end{subequations}
\end{widetext}
\endgroup

%%% Local Variables:
%%% mode: latex
%%% TeX-master: "paper1"
%%% End:

Some of these terms were present in the mode decompositions given in
Ref.~\cite{arunetal:2009}.  By taking $\iota = 0$, $\Psi=0$, and
$\alpha = -\pi/2$ in that reference, and noting that their conventions
for the polarization tensor give rise to relative factors of
$(-\i)^{m}$, we find agreement with all of the corresponding terms
given here.

%%%%%%%%%%%%%%%%%%%%%%%%%%%%%%%%%%%%%%%%%%%%%%%%%%%%%%%%%%%%%%%%%%%%%%
%%%%%%%%%%%%%%%%%%%%%%%%%%%%%%%%%%%%%%%%%%%%%%%%%%%%%%%%%%%%%%%%%%%%%%

\section{Efficient calculation of waveforms from rotating frames for
  data analysis}
\label{sec:WaveformsForDataAnalysis}

As mentioned previously, it is a simple matter to transform the
waveform between frames using Eq.~\eqref{eq:ModeRotation}.  For
example, we might transform the waveform $\Rotated{h}$ from the
co-orbital frame back into $h$ as seen in the inertial frame, and then
evaluate the waveform at a point, to give us the data that would be
measured by a gravitational-wave detector.  But this is very
computationally expensive if the end goal is simply to obtain the
waveform along a single world line.  Even if we only include the
$\ell=2$ component, rotation of the waveform requires calculating all
$25$ elements of the $\D^{(2)}$ matrix; including through $\ell=8$
would require calculating $959$ elements of the various $\D^{(\ell)}$
matrices.  And because the rotation $\rotor{R}$ changes from instant
to instant, these calculations would all need to be redone at each
time step.  The waveform $h$ in the inertial from would then be
evaluated as
\begin{equation}
  \label{eq:EvaluateWaveformInInertialFrame}
  h(\vartheta, \varphi) = \sum_{\ell,m} h^{\ell,m}\,
  \mTwoYlm{\ell,m}(\vartheta, \varphi).
\end{equation}
The SWSH components in this expression would only need to be
calculated once per waveform, and there are only $5$ to evaluate for
the $\ell=2$ component, or $77$ when including through $\ell=8$, so
this would be a very small portion of the computation.

Fortunately, there is a far better alternative.  We can evaluate the
waveform as given in the co-orbital (or any other) frame, but change
the point at which the evaluation takes place, to cancel out the
rotation.  This would require no evaluations of $\D^{(\ell)}$
matrices, but $5$ ($77$) evaluations of the SWSH components at each
time step when including all modes through $\ell=2$ ($\ell=8$).  This
is a substantial savings, and could easily be implemented using
existing software packages---for example, the \software{spinsfast}
package~\cite{huffenbergerwandelt:2010} developed for the
cosmic-microwave-background community, or the
\software{SphericalFunctions} module included in the
\href{\ancillaryURL} {ancillary files} on this paper's arXiv page.

One subtlety must be handled carefully.  We would write the waveform
in the inertial frame as
\begin{equation}
  \label{eq:EvaluateWaveformInRotatingFrame}
  h(\vartheta, \varphi) = \sum_{\ell,m} \Rotated{h}^{\ell,m}\,
  \mTwoYlm{\ell,m}(\Rotated{\vartheta}, \Rotated{\varphi})\, \e^{2\,
    \i\, \Rotated{\gamma}},
\end{equation}
where not only $(\Rotated{\vartheta}, \Rotated{\varphi})$ must be
computed, but also the angle $\Rotated{\gamma}$.  The latter is
necessary because we must evaluate the spin-weighted field with
respect to the tangent basis that would be used at the inertial point
$(\vartheta, \varphi)$.  But the tangent basis of the
$\mTwoYlm{\ell,m}$ function is defined naively, with respect to the
input arguments $(\Rotated{\vartheta}, \Rotated{\varphi})$, and would
therefore change from moment to moment in ways completely unrelated to
the inertial tangent basis.\footnote{Another way of saying this is to
  note that, unlike the SWSH \emph{modes} $h^{\ell,m}$, the SWSHs
  \emph{functions} $\mTwoYlm{\ell,m}$ do not transform among
  themselves under rotations~\cite{boyle:2013}; only the full
  $\D^{(\ell)}_{m',m}$ functions do that, which is why we must use
  them in what follows.}  Unless we account for $\Rotated{\gamma}$,
our predicted waveform would be different from the physical waveform
by an arbitrary and erratically time-dependent
phase~\cite{boyleetal:2011}---equivalent to a wildly rotating
detector.

We can evaluate the right-hand side above, using the expression for
SWSHs in terms of the Wigner $\D$ matrices:
\begin{equation}
  \label{eq:SWSHsAsWignerDs}
  \sYlm{\ell,m}(\vartheta, \varphi) = (-1)^{s} \sqrt{\frac{2\ell+1}
    {4\pi}}\, \D^{(\ell)}_{m,-s} (\rotor{R}_{\vartheta,\varphi}),
\end{equation}
where $\rotor{R}_{\vartheta, \varphi}$ rotates $\zHat$ onto the point
with coordinates $(\vartheta, \varphi)$.  Now, combining this with
Eqs.~\eqref{eq:ModeRotation}
and~\eqref{eq:EvaluateWaveformInInertialFrame}, we can calculate
\begin{subequations}
  \label{eq:EvaluateWaveformByRotations}
  \begin{align}
    h(\vartheta, \varphi) &= \sum_{\ell,m}
    h^{\ell,m}\, \mTwoYlm{\ell,m}(\vartheta, \varphi) \\
    &= \sum_{\ell,m} \sum_{m'} \Rotated{h}^{\ell,m'}\,
    \D^{(\ell)}_{m',m}(\Rf^{-1})\, \sqrt{\frac{2\ell+1} {4\pi}}\,
    \D^{(\ell)}_{m,2}
    (\rotor{R}_{\vartheta,\varphi}) \\
    \label{eq:EvaluateWaveformByRotations_D}
    &= \sum_{\ell,m} \Rotated{h}^{\ell,m}\, \sqrt{\frac{2\ell+1}
      {4\pi}}\, \D^{(\ell)}_{m,2} (\Rf^{-1}\,
    \rotor{R}_{\vartheta,\varphi}).
  \end{align}
\end{subequations}
Here, $\Rf$ is the rotation that takes the inertial frame onto the
rotating frame, and we have used the composition property of the
$\D^{(\ell)}$ matrices to combine two into one.

We emphasize that Eq.~\eqref{eq:EvaluateWaveformByRotations_D} is the
preferred way to evaluate $h(\vartheta, \varphi)$, since evaluation of
the general SWSH is essentially the same as evaluating elements of the
Wigner $\D^{(\ell)}$ matrix, in terms of computational cost.
Moreover, the \software{SphericalFunctions} module implements
efficient and stable evaluation of the $\D^{(\ell)}$ matrices directly
in terms of quaternions, meaning that the product $\Rf^{-1}\,
\rotor{R}_{\vartheta,\varphi}$ can be evaluated by simple quaternion
multiplication.  The accompanying \software{GWFrames} module uses this
technique for evaluating rotating-frame waveforms at a point.

Although Eq.~\eqref{eq:EvaluateWaveformByRotations_D} is a far
superior formula for calculating $h(\vartheta, \varphi)$, it may also
be useful for testing purposes or for working with older software
libraries to rewrite the result in precisely the terms of
Eq.~\eqref{eq:EvaluateWaveformInRotatingFrame}.  We caution, however,
that this is tantamount to using Euler angles.  To be precise,
$(\Rotated{\varphi}, \Rotated{\vartheta}, \Rotated{\gamma})$ [note the
ordering] is the Euler-angle form of the rotation $\Rf^{-1}\,
\rotor{R}_{\vartheta,\varphi}$ under the $z$-$y$-$z$
convention~\cite{boyle:2013}.  Of course, the Euler-angle
representation of rotations is dramatically inferior to quaternions in
almost every way---in this case, because composing rotations given in
terms of Euler angles is computationally expensive and inaccurate
(essentially requiring conversion to another form and back), and
because the $\D^{(\ell)}$ matrices can be calculated more accurately
and efficiently using quaternions directly.  Nonetheless, we can at
least reduce one obstacle on this path by showing how to derive the
angles after composing the rotations by quaternion multiplication.

We need to calculate $\Rotated{\vartheta}$, $\Rotated{\varphi}$, and
$\Rotated{\gamma}$ such that
\begin{equation}
  \label{eq:RotationsRelation}
  \Rf^{-1}\, \rotor{R}_{\vartheta,\varphi} =
  \rotor{R}_{\Rotated{\vartheta}, \Rotated{\varphi}}\,
  \e^{\Rotated{\gamma}\, \zHat/2}.
\end{equation}
In the last term on the right-hand side of this equation, we have used
quaternion notation to express a rotation through $\Rotated{\gamma}$
about the $z$ axis.\footnote{Essentially, $\zHat$ is the generator of
  rotations about the $z$ axis, and a unit quaternion is equivalent to
  the square-root of the usual rotation matrix---hence the factor of
  $1/2$.  For a more complete explanation of quaternions as applied to
  rotations, see Sec.~I~C or Appendix~A of Ref.~\cite{boyle:2013}.}
Now, we know both quantities on the left-hand side of the equation, so
we can simply evaluate their product, and define the quaternion
components of the result as $(r_{0}, r_{1}, r_{2}, r_{3})$.
Straightforward calculation shows that our angles are given by the
following simple formulas:
\begin{subequations}
  \label{eq:ConvertRotorToEulerAngles}
  \begin{align}
    \label{eq:ConvertRotorToEulerAngles_theta}
    \Rotated{\vartheta} &= 2 \arccos \sqrt{r_{0}^{2} + r_{3}^{2}}, \\
    \label{eq:ConvertRotorToEulerAngles_phi}
    \Rotated{\varphi} &= \arctan \frac{r_{3}} {r_{0}}
    + \arctan \frac{-r_{1}} {r_{2}}, \\
    \label{eq:ConvertRotorToEulerAngles_gamma}
    \Rotated{\gamma} &= \arctan \frac{r_{3}} {r_{0}} - \arctan
    \frac{-r_{1}} {r_{2}}.
  \end{align}
\end{subequations}
Using these results, Eq.~\eqref{eq:EvaluateWaveformByRotations}
becomes
\begin{subequations}
  \label{eq:EvaluateWaveformByRotations2}
  \begin{align}
    h(\vartheta, \varphi) &= \sum_{\ell,m} \Rotated{h}^{\ell,m}\,
    \sqrt{\frac{2\ell+1} {4\pi}}\, \D^{(\ell)}_{m,2}
    (\rotor{R}_{\Rotated{\vartheta}, \Rotated{\varphi}}\,
    \e^{\Rotated{\gamma}\, \zHat/2}) \\
    &= \sum_{\ell,m} \Rotated{h}^{\ell,m}\, \sqrt{\frac{2\ell+1}
      {4\pi}}\, \D^{(\ell)}_{m,2} (\rotor{R}_{\Rotated{\vartheta},
      \Rotated{\varphi}})\, \,
    \e^{2\, \i\, \Rotated{\gamma}} \\
    \label{eq:EvaluateWaveformInRotatingFrame_SWSH}
    &= \e^{2\, \i\, \Rotated{\gamma}}\, \sum_{\ell,m}
    \Rotated{h}^{\ell,m}\, \mTwoYlm{\ell,m}(\Rotated{\vartheta},
    \Rotated{\varphi}).
  \end{align}
\end{subequations}
In going from the first to the second line, we have used the
composition property of the $\D^{(\ell)}$ matrices, and the simple
expression $\D^{(\ell)}_{m',2}(\e^{\Rotated{\gamma}\zHat/2}) =
\e^{2\i\Rotated{\gamma}} \delta_{m',2}$.  The sum in
Eq.~\eqref{eq:EvaluateWaveformInRotatingFrame_SWSH} is precisely the
naive evaluation of the waveform in the rotating frame, which may be
readily available from certain software packages.

Again, we discourage the use of
Eq.~\eqref{eq:EvaluateWaveformInRotatingFrame_SWSH}, as it requires
the additional intermediate calculations of
Eqs.~\eqref{eq:ConvertRotorToEulerAngles}, along with the usual
numerical inaccuracies and edge cases to deal with; we advise using
Eq.~\eqref{eq:EvaluateWaveformByRotations_D} instead, with direct
evaluation of the $\D^{(\ell)}$ matrices in terms of quaternions.
However, either expression eliminates the need to rotate the waveform
modes themselves, which reduces the number of $\D^{(\ell)}$ elements
needed by an order of magnitude at each time step.  Because the
complete expressions for the \pN waveform are nearly as simple as the
expressions for nonprecessing systems, and because the system's
orbital trajectories ($\Rf$, here) needed to be calculated in any
case, this gives us an efficient method for producing an accurate
waveform for use in data analysis.  While this will presumably still
be computationally expensive compared to using simple quadrupolar
waveforms, whenever higher modes or generally greater accuracy are
needed, this approach can be faster and more accurate than using
expressions for waveforms in which the rotation is explicitly
included.

%%%%%%%%%%%%%%%%%%%%%%%%%%%%%%%%%%%%%%%%%%%%%%%%%%%%%%%%%%%%%%%%%%%%%%
%%%%%%%%%%%%%%%%%%%%%%%%%%%%%%%%%%%%%%%%%%%%%%%%%%%%%%%%%%%%%%%%%%%%%%
\section{Parity and antipodes}
\label{sec:Parityandantipodes}

It is vital to understand in detail the behavior of waveforms and---in
particular---the SWSH modes of waveforms under parity conjugation and
evaluation at antipodes.  Here, we describe these details in greater
generality than is necessary for our purposes in the main body of this
paper, in the hope that this more complete view will clarify the
issues involved.

We begin by defining the three parity operators
$\ParityInversion_{x}$, $\ParityInversion_{y}$, and
$\ParityInversion_{z}$, which represent reflections along the given
axes.  So, for example, $\ParityInversion_{z}$ represents reflection
across the $x$-$y$ plane; vectors under this operation just get a sign
flip in the $z$ component.  We denote by $\ParityInversion_{-}$ the
more standard parity operation in three-dimensional physics, which is
reversal of all components of a vector.  Equivalently,
$\ParityInversion_{-}$ is the composition of the previous three
operations, or any one of them followed by rotation through $\pi$
about that same axis.  For simplicity, we denote any one of these four
operators as $\ParityInversion_{i}$.

For a spin-zero field $f$, its parity-conjugate field is found by
simply evaluating the original field at the parity-conjugate location.
Thus, for any direction $\rHat$, we can relate the values of the field
$f$ and its parity-conjugate field $\ParityInversion_{i}\{f\}$ as
\begin{equation}
  \label{eq:ParityForSpinZero}
  \ParityInversion_{i}\{f\}\left( \rHat \right) = f
  \left( \ParityInversion_{i} \left\{\rHat\right\} \right).
\end{equation}
In this sense, spin-zero fields transform as ``true scalars''.

Spin-\emph{weighted} fields, on the other hand, do not transform so
simply, because they are defined with respect to a coordinate basis
for the tangent space of the sphere, and that basis is also affected
by the parity operation.  In terms of the standard $(\hat{\vartheta},
\hat{\varphi})$ basis, we have the following transformations:
\begin{subequations}
  \begin{align}
    \ParityInversion_x &: \hat{\vartheta}(\rHat) \mapsto
    \hat{\vartheta}(\ParityInversion_x\rHat)& \ParityInversion_x &:
    \hat{\varphi}(\rHat) \mapsto
    -\hat{\varphi}(\ParityInversion_x\rHat),
    \\
    \ParityInversion_y &: \hat{\vartheta}(\rHat) \mapsto
    \hat{\vartheta}(\ParityInversion_y\rHat)& \ParityInversion_y &:
    \hat{\varphi}(\rHat) \mapsto
    -\hat{\varphi}(\ParityInversion_y\rHat),
    \\
    \ParityInversion_z &: \hat{\vartheta}(\rHat)\mapsto
    -\hat{\vartheta}(\ParityInversion_z\rHat)& \ParityInversion_z &:
    \hat{\varphi}(\rHat) \mapsto
    \hat{\varphi}(\ParityInversion_z\rHat),
    \\
    \ParityInversion_- &: \hat{\vartheta}(\rHat)\mapsto
    -\hat{\vartheta}(\ParityInversion_-\rHat)& \ParityInversion_- &:
    \hat{\varphi}(\rHat) \mapsto
    \hat{\varphi}(\ParityInversion_-\rHat).
  \end{align}
\end{subequations}
Thus, the usual complex basis representation $m_{j} \defined
\vartheta_{j} + i\, \varphi_{j}$ transforms under these parity
operations as
\begin{subequations}
  \begin{align}
    \ParityInversion_{x}, \ParityInversion_{y} &: m_{j} \mapsto
    \bar{m}_{j},
    \\
    \ParityInversion_{z}, \ParityInversion_{-} &: m_{j} \mapsto
    -\bar{m}_{j}.
  \end{align}
\end{subequations}
Since $h$ has spin weight $s=-2$, it is defined by contraction of a
tensor field with \emph{two} copies of $\bar{m}$, parity conjugation
simply induces complex conjugation of the field:
\begin{equation}
  \label{eq:ParityForSpinTwo}
  \ParityInversion_{i}\{h\}\left( \rHat \right) = \bar{h}
  \left( \ParityInversion_{i} \left\{\rHat\right\} \right).
\end{equation}
This is distinct from the transformation of spin-zero fields
\eqref{eq:ParityForSpinZero} only by virtue of the complex conjugation
on the right-hand side.  It is worth noting that similar expressions
for spin weight $s=-1$ fields, for example, would also involve factors
of $-1$ for $\ParityInversion_{z}$ and $\ParityInversion_{-}$.

We can also distinguish the parity operations from a similar
operation: evaluation at the antipode, which we denote by $A$.  The
distinction is important because the tangent basis used for evaluation
at the antipode is just the standard basis at that point; the vectors
are not affected by the operation $A$, as they are for the
$\ParityInversion_{i}$ operations.  Thus,
\begin{equation}
  \label{eq:AntipodesForSpinTwo}
  A\{f\}\left( \rHat \right) = f \left( A \left\{\rHat\right\} \right)
  = f \left( -\rHat \right),
\end{equation}
for a field of \emph{any} spin weight (or even ill defined spin
weight).  Unfortunately, because of the differential behavior of
$-\rHat$, this produces a field with the opposite spin.  Thus, a more
convenient operator is $\ConjugateAntipodalEvaluation$, which also
conjugates the field.  For the special case of operations on $h$, it
is clear that $\ConjugateAntipodalEvaluation =
\ParityInversion_{-}$:
\begin{equation}
  \label{eq:EqualityOfABarAndPMinus}
  \ConjugateAntipodalEvaluation\{h\} = \ParityInversion_{-}\{h\}.
\end{equation}
However, we regard this relation as mere coincidence, and consider
$\ConjugateAntipodalEvaluation$ to be the more interesting operator in
general.

It is convenient to find simple expressions for the behavior of modes
under each of our five transformations.  The main task is to find
expressions for the SWSHs under evaluation at the various conjugated
points.  This is easily done for $s=-2$ by inspection of the
functions:
\begin{subequations}
  \label{eq:SWSHParities}
  \begin{gather}
    \sYlm{\ell,m} \left( \ParityInversion_{x}\rHat \right) =
    (-1)^{m}\,
    \sYbarlm{\ell,m} \left( \rHat \right), \\
    \sYlm{\ell,m} \left( \ParityInversion_{y}\rHat \right) =
    \sYbarlm{\ell,m} \left( \rHat \right), \\
    \sYlm{\ell,m} \left( \ParityInversion_{z}\rHat \right) =
    (-1)^{\ell+s}
    \sYbarlm{\ell,-m} \left( \rHat \right), \\
    \sYlm{\ell,m} \left( -\rHat \right) = (-1)^{\ell+s+m}
    \sYbarlm{\ell,-m} \left( \rHat \right).
  \end{gather}
\end{subequations}
Now, for example, we can write
\begin{subequations}
  \begin{align}
    \label{eq:Pzh}
    \ParityInversion_{z}\{h\}\left(\rHat\right) &=
    \bar{h}\left(\ParityInversion_{z} \rHat \right)
    \\
    &= \sum_{\ell,m} \bar{h}^{\ell,m} \mTwoYbarlm{\ell,m}
    \left(\ParityInversion_{z} \rHat \right)
    \\
    &= \sum_{\ell,m} \bar{h}^{\ell,m} (-1)^{\ell} \mTwoYlm{\ell,-m}
    \left( \rHat \right)
    \\
    &= \sum_{\ell,m} (-1)^{\ell} \bar{h}^{\ell,-m} \mTwoYlm{\ell,m}
    \left( \rHat \right)
    \\
    &= \sum_{\ell,m} \ParityInversion_{z}\{h\}^{\ell,m}
    \mTwoYlm{\ell,m} \left( \rHat \right),
  \end{align}
\end{subequations}
where the modes of the parity-conjugate field are related to the
original modes by
\begin{equation}
  \label{eq:PzhModes}
  \ParityInversion_{z}\{h\}^{\ell,m} = (-1)^{\ell} \bar{h}^{\ell,-m}.
\end{equation}
Finally, we can see that Eq.~\eqref{eq:ModeSymmetry} is really the
statement that the modes are invariant under reflection across the
$x$-$y$ plane.  That is, Eq.~\eqref{eq:ModeSymmetry} is just the
statement $\ParityInversion_{z}\{h\}^{\ell,m} = h^{\ell,m}$, which is
itself a specific example of the fact that nonprecessing systems
themselves are invariant under that reflection.

For completeness, we now list the transformation laws for the modes
under each of the operations (which we note are only valid for fields
of spin weight $s=-2$).
\begin{subequations}
  \label{eq:ModeTransformationLaws}
  \begin{align}
    \ParityInversion_{x}\{h\}^{\ell,m} &= (-1)^{m} \bar{h}^{\ell,m}, \\
    \ParityInversion_{y}\{h\}^{\ell,m} &= \bar{h}^{\ell,m}, \\
    \ParityInversion_{z}\{h\}^{\ell,m} &= (-1)^{\ell} \bar{h}^{\ell,-m}, \\
    \ParityInversion_{-}\{h\}^{\ell,m} =
    \ConjugateAntipodalEvaluation\{h\}^{\ell,m} &= (-1)^{\ell+m}
    \bar{h}^{\ell,-m}.
  \end{align}
\end{subequations}
The details of these equations are important.  In particular, they
show that $\ParityInversion_{x}$, $\ParityInversion_{y}$, and
$\ParityInversion_{z}$ do not behave well under rotation; rotation of
the parity-conjugated fields is not the same as parity conjugation of
the rotated fields.  This is unsurprising, because those parity
operators are defined with respect to basis vectors, which are
obviously not rotationally invariant; the operators do not commute
with the usual angular-momentum operator $\threevec{L}$.  On the other
hand, $\ParityInversion_{-}$ and $\ConjugateAntipodalEvaluation$
\emph{do} behave well under rotations.  For this reason, we can define
rotationally invariant quantities using, for example, $\int
\abs{h-\ConjugateAntipodalEvaluation\{h\}}^{2} d\Omega$, whereas we
need to optimize over attitude when defining similar quantities with
respect to $\ParityInversion_{z}$, for example.  We use these
properties in Sec.~\ref{sec:Invariantcombinationsofmodes}.

%%%%%%%%%%%%%%%%%%%%%%%%%%%%%%%%%%%%%%%%%%%%%%%%%%%%%%%%%%%%%%%%%%%%%%
%%%%%%%%%%%%%%%%%%%%%%%%%%%%%%%%%%%%%%%%%%%%%%%%%%%%%%%%%%%%%%%%%%%%%%
%% References

%% Try this if the last two columns before the bib don't break nicely
% \vspace{0.1in}

\vfil

%% Try this if missing footnote on page with references:
% \clearpage

%% Bibtex sometimes uses these automatically, so we have to make sure
%% they have their original meanings when it does
\let\c\Originalcdefinition \let\d\Originalddefinition
\let\i\Originalidefinition

%% Include any .bib files that need to be referenced
\bibliography{References_NumericalRelativity,Infrastructure/References_AnalyticalWaveforms,Infrastructure/References_GeometricAlgebra,Infrastructure/References_Detection,Infrastructure/References_ParameterEstimation,Infrastructure/References_General,Infrastructure/References_Astro,Infrastructure/References_Scri}

%%%%%%%%%%%%%%

\end{document}